\documentstyle[preprint,aps]{revtex}
\begin{document}
\title{Nonlinear Connections in Superbundles \\
and Locally Anisotropic Supergravity}
\newtheorem{theorem}{Theorem}
\newtheorem{defin}{Definition}
\draft
\author{Sergiu I. Vacaru}
\address{Institute of Applied Physics, Academy of Sciences,\\
5 Academy str., Chi\c{s}in\u{a}u--028, Moldova}
\date{\today}
\maketitle
\begin{abstract}
The theory of locally anisotropic superspaces (supersymmetric generalizations
of various types of Kaluza--Klein, Lagrange and Finsler spaces) is laid down.
 In this framework we perform
the analysis of  construction  of the supervector bundles provided with
nonlinear and distinguished connections and metric structures. Two models of
locally anisotropic  supergravity are proposed and studied in details.

{\bf \copyright S.I.Vacaru}
\end{abstract}

\widetext
\section{Introduction}
\label{sec:level1}

Differential geometric techniques plays an important role in formulation
and mathematical formalization of models of fundamental interactions of
physical fields. In the last twenty years there has been a substantial
interest in the construction of differential supergeometry with the aim
of getting a framework for the supersymmetric field theories (the theory
of graded manifolds [1-4] and the theory of supermanifolds [5-9]). Detailed
considerations of geometric and topological aspects of supermanifolds and
formulation of superanalysis are  contained in [10-16].
\par
 Spaces with local anisotropy are used in some divisions of
theoretical and mathematical physics [17-20] (recent applications in physics
and biology are summarized in [21,22]). The first models of locally anisotropic (la)
spaces (la--spaces) have been proposed by P.Finsler [23] and E.Cartan [24].
 Early approaches and
modern treatments of Finsler geometry and its extensions can be found in
[25-30]. We shall use the general approach to the geometry of la--spaces,
developed by R.Miron and M.Anastasiei [26,27], as a starting point for our
definition of superspaces with local anisotropy and formulation of la--supergravitational
models.
\par
 In different models of la--spaces one considers nonlinear and linear
connections and metric structures in vector and tangent bundles on locally
isotropic space--times ((pseudo)--Riemannian, Einstein--Cartan and more general
types of curved spaces with torsion and nonmetricity). It seems likely that
la--spaces make up a more convenient geometric background for developing in a
selfconsistent manner classical and quantum statistical and field theories
in non homogeneous, dispersive media with radiational, turbulent and random
processes.In [31-35] some variants of Yang--Mills, gauge gravity and the definition
of spinors on la--spaces have been proposed. In connection with the above
mentioned the formulation of supersymmetric extensions of classical and
quantum field theories on la--spaces presents
 a certain interest
\par
In works [36--38] a new viewpoint on differential geometry of supermanifolds
is discussed. The author introduced the nonlinear connection (N--connection)
structure and developed  a corresponding distinguished by N--connection
supertensor covariant differential calculus in the frame of  De Witt [5]
approach to supermanifolds, by considering the particular case of superbundles
with typical fibres parametrized by noncommutative coordinates. This is the
first example of superspace with local anisotropy. But up to the present
we have not a general, rigorous mathematical, definition of locally anisotropic
superspaces (la--superspaces).
\par
In this paper we intend to give some contributions to the theory of vector and
tangent superbundles provided with nonlinear and distinguished connections and
metric structures (a generalized model of la--superspaces). Such superbundles
contain as particular cases the supersymmetric extensions of Lagrange and
Finsler spaces. We shall also formulate and analyze two models of locally
anisotropic supergravity.
\par
The plan of the work is the following: After giving in Sec. II the basic terminology
on supermanifolds and superbundles, in Sec.III we introduce nonlinear and linear
distinguished connections in vector superbundles.The geometry of the total space
of vector superbundles will be studied in Sec.IV by considering distinguished
connections and their structure equations. Generalized Lagrange and Finsler
superspaces will be defined in Sec.V. In Sec.VI the Einstein equations on the
la--superspaces are written and analyzed. A version of gauge like la--supergravity
will be also proposed. Concluding remarks and discussion are contained in Sec.VII.

\section{Supermanifolds and Superbundles}

In this section we outline some necessary definitions, concepts and results
on the theory of supermanifolds (s--manifolds) [5--14].
\par
The basic structures for building up s--manifolds (see [6,9,14]) are \ Grassmann
algebra and Banach space. Grassmann algebra is considered a real associative
algebra $\Lambda$ (with unity) possessing a finite (canonical) set of anticommutative
generators $\beta_{\hat A}$, ${{[{\beta_{\hat A}},{\beta_{\hat B}}]}_{+}}=
{{\beta_{\hat A}}}{{\beta_{\hat C}}}+ {{\beta_{\hat C}}}{{\beta_{\hat A}}}=0$,
where ${{\hat A},{\hat B},...}=1,2,...,{\hat L}$. This way it is defined a
 ${Z_2}$-graded
commutative algebra ${{\Lambda}_0}+{{\Lambda}_1}$,
 whose even part ${{\Lambda}_0}$
(odd part ${{\Lambda}_1}$) represents a ${2^{{\hat L}-1}}$--dimensional real
 vector space
of even (odd) products of generators ${\beta}_{\hat A}$.After setting
${{\Lambda}_0}={\cal R}+{{\Lambda}_0}'$, where ${\cal R}$   is the real
number field and ${{\Lambda}_0}'$ is the subspace of ${\Lambda}$ consisting
of nilpotent elements, the projections ${\sigma}: {\Lambda}\to{\cal R}$ and
$s: {\Lambda}\to{{\Lambda}_0}'$ are called, respectively, the body and soul maps.
\par
A Grassmann algebra can be provided with both structures of a Banach algebra
and Euclidean topological space by the norm [6]
$${\Vert}{\xi}{\Vert}={{\Sigma}_{{\hat A}_i}}{\vert}a^{{{\hat A}_1}...{{\hat A}_k}}
{\vert}, {\xi}={{\Sigma}_{r=0}^{\hat L}} a^{{{\hat A}_1}...{{\hat A}_r}}
{{\beta}_{{\hat A}_1}}...{{\beta}_{{\hat A}_r}}.$$
A superspace is defined as a product
$$ {\Lambda}^{n,k}={\underbrace{{{\Lambda}_0}{\times}...{\times}
{{\Lambda}_0}}_{n} {\times}
{\underbrace{{{\Lambda}_1}{\times}...{\times}
{{\Lambda}_1}}_{k}}}{\quad}.$$
This represents the $\Lambda$-envelope of a $Z_2$-graded vector space ${V^{n,k}}=
{V_0}{\otimes}{V_1}={{\cal R}^n}\oplus{{\cal R}^k}$, which is obtained by multiplication
of even (odd) vectors of $V$ by even (odd) elements of ${\Lambda}$. The superspace
(as the ${\Lambda}$-envelope) posses $(n+k)$ basis vectors
$\{ {\hat {{\beta}_i}},
{\quad}
  i=0,1,...,n-1,$ and ${\quad}{{\beta}_{\hat i}}, {\quad},{\hat i}=1,2,...k \}$.
Coordinates of even (odd) elements of $V^{n,k}$ are even (odd) elements of
$\Lambda$. On the other hand, a superspace $V^{n,k}$ forms a $({2^{{\hat L}-1}})
(n+k)$-dimensional real vector spaces with a basis
$\{{{\hat {\beta}}_{i({\Lambda})}},{{\beta}_{{\hat i}({\Lambda})}}\}$.
\par
Functions of superspaces, differentiation with respect to Grassmann coordinates
,supersmooth (superanalytic) functions and mappings are defined by analogy with
the ordinary case, but with a glance to certain specificity caused by changing
of real (or complex) number field into Grassmann algebra $\Lambda$. Here we
remark that functions on a superspace ${\Lambda}^{n,k}$ which takes values
in Grassmann algebra can be considered as mappings of the space ${\cal R}^
{{({2^{({\hat L}-1)}})}{(n+k)}}$ into the space ${\cal R}^{2{\hat L}}$. Functions
being differentiable with regard to Grassmann coordinates can be rewritten via
derivatives on real coordinates, which obey a generalized
version of Cauchy-Riemann conditions.
\par
A $(n,k)$-dimensional s-manifold $M$ is defined as a Banach manifold (see,
for example, [39]) modelled on ${\Lambda}^{n,k}$ endowed with an atlas
${\psi}={\lbrace}{U_{(i)}},{{\psi}_{(i)}}:{U_{(i)}}\to{{\Lambda}^{n,k}}, (i)\in J
{\rbrace}$ whose transition functions ${\psi}_{(i)}$ are supersmooth [6,9].
Instead of supersmooth functions we can use $G^{\infty}$-functions [6] and
define $G^{\infty}$-supermanifolds ($G^{\infty}$ denotes the class of superdifferentiable
functions). The local structure of a $G^{\infty}$-supermanifold can be built
very much as on a $C^{\infty}$-manifold. Just as a vector field on a $n$-dimensional
$C^{\infty}$-manifold can be expressed locally as
$${\Sigma}^{n-1}_{i=0}{\quad}{f_i}{(x^j)}{{\partial}\over{{\partial}x^i}},$$
where $f_i$ are $C^{\infty}$-functions, a vector field on an $(n,k)$-dimensional
$G^{\infty}$-supermanifold $M$ can be expressed locally on an open region
$U{\subset}M$ as $$
{\Sigma}^{n-1+k}_{I=0}{\quad}{f_I}{(x^J)}{{\partial}\over{{\partial}x^I}}=$$
$${\Sigma}^{n-1}_{i=0}{\quad}{f_i}{(x^{j},{{\theta}^{\hat j}})}{{\partial}\over
{{\partial}x^i}}+
{\Sigma}^{k}_{{\hat i}=1}{\quad}{f_{\hat i}}{(x^{j},{{\theta}^{\hat j}})}
{\partial\over\partial {{\theta}^{\hat i}}}, $$
where $x=({\hat x},{\theta})=\lbrace {x^I}=({{\hat x}^i},{\theta}^{\hat i})
\rbrace$ are local (even, odd) coordinates. We shall use indices $I=(i,{\hat i}),
J=(j,{\hat j}),K=(k,{\hat k}),...$ for geometric objects on $M$. A vector field
on $U$ is an element $X{\subset}End[{G^{\infty}}(U)]$ (we can also consider
supersmooth functions instead of $G^{\infty}$-functions) such that
$$ X(fg)=(Xf)g+{(-)}^{{\mid}f{\mid}{\mid}X{\mid}} fXg,$$
for all $f,g$ in $G^{\infty}(U)$, and
$$X(af)={(-)}^{{\mid}X{\mid}{\mid}a{\mid}}aXf,$$
where ${\mid}X{\mid}$ and ${\mid}a{\mid}$ denote correspondingly the parity
$(=0,1)$ of values $X$ and $a$ and for simplicity in this work we shall write
${(-)}^{{\mid}f{\mid}{\mid}X{\mid}}$ instead of ${(-1)}^{{\mid}f{\mid}{\mid}X{\mid}}.$
\par
A super Lie group (sl-group) [7] is both an abstract group and a s-manifold,
provided that the group composition law fulfils a suitable smoothness
condition (i.e. to be superanalytic, for short,$sa$ [9]).
\par
In our further considerations we shall use the group of automorphisms of
${\Lambda}^{(n,k)}$, denoted as $GL(n,k,{\Lambda})$, which can be parametrized
as the super Lie group of invertible matrices
$$Q={\left(\begin{array}{cc} A & B \\
                            C & D \end{array} \right) } ,$$
where A and D are respectively $(n{\times}n)$ and $(k{\times}k)$
matrices consisting of even Grassmann elements and B and C are rectangular
matrices consisting of odd Grassmann elements. A matrix Q is invertible as soon
as maps ${\sigma}A$ and ${\sigma}D$ are invertible matrices.A sl-group
represents an ordinary Lie group included in the group of linear transforms
$GL(2^{{\hat L}-1}(n+k),{\cal R})$. For matrices of type Q one defines [1-3]
the superdeterminant,$sdetQ$, supertrace, $strQ$, and superrank,$srankQ$.
\par
One calls Lie superalgebra (sl-algebra) any $Z_2$-graded algebra $A={A_0}\oplus
A_1$ endowed with product $[,\}$ satisfying the following properties:
$$[I,I'\}=-{(-)}^{{\mid}I{\mid}{\mid}I'{\mid}}[I',I\},$$
$$[I,[I',I''\}\}=[[I,I'\},I''\}+{(-)}^{{\mid}I{\mid}{\mid}I'{\mid}}[I'[I,I''\}\},$$
$I{\in}A_{{\mid}I{\mid}}, {\quad} I'{\in}A_{{\mid}I'{\mid}}$, where ${\mid}I{\mid},
{\mid}I'{\mid}=0,1$ enumerates, respectively, the possible parity of elements
$I,I'$. The even part $A_0$ of a sl-algebra is a usual Lie algebra and the odd
part $A_1$ is a representation of this Lie algebra.This enables us to classify
sl--algebras following the Lie algebra classification [40]. We also point out
that irreducible linear representations of Lie superalgebra A are realized in
$Z_2$-graded vector spaces by matrices
$\left( \begin{array}{cc} A & 0 \\
                          0 & D \end{array} \right)$ for even elements and
$\left( \begin{array}{cc} 0 & B \\
                          C & 0  \end{array} \right)$
for odd elements and that, roughly speaking, A is a superalgebra of generators
of a sl-group.
\par
An sl--module $W$ (graded Lie module) [7] is a $Z_2$-graded left $\Lambda$-module
endowed with a product $[,\}$ which satisfies the graded Jacobi identity and makes
$W$ into a graded-anticommutative Banach algebra over $\Lambda$.
 One calls the Lie module
{\cal G} the set of the left-invariant derivatives of a sl-group $G$.
\par
One constructs the supertangent bundle (st-bundle) $TM$ over a s-manifold $M$,
${\pi}:TM\to{M}$ in a usual manner (see, for instance,[39]) by taking as the
typical fibre the superspace ${\Lambda}^{n,k}$ and as the structure group
the group of automorphisms, i.e. the sl-group $GL(n,k,{\Lambda}).$
\par
A s-manifold and a st-bundle $TM$ may be represented as a certain $2^{{\hat L}-1}
(n+k)$-dimensional real manifold and the tangent bundle over it whose transition
function obey the special conditions of Cauchy-Riemann type.
\par
Let us denote ${\hat {\cal E}}$ a vector superspace (vs-space) of dimension $(m,l)$
(with respect to a chosen base we parametrize an element $y\in{\hat {\cal E}}$ as
$y=({\hat y}, \zeta )=\{{y^A}=({\hat {y^a}},{\zeta}^{\hat a})\}$, where
$a=1,2,...,m $ and ${\hat a} =1,2,...,l$). We shall use indices
$A=(a,{\hat a}), B=(b,{\hat b}),...$ for objects on vs-spaces. A vector superbundle
(vs-bundle) ${\cal E}$ over base $M$ with total superspace $E$, standard fibre
${\hat {\cal F}}$ and surjective projection ${{\pi}_E}:E{\to}M$ is defined (see
details and variants in [11,16]) as in the case of ordinary manifolds (see,
for instance, [39,26,27]). A section of $\cal E$ is a supersmooth map $s:U{\to}E$
such that ${{\pi}_E}{\cdot}s=id_U.$
\par
A subbundle of ${\hat {\cal E}}$ is a triple $(B,f,f')$, where $B$ is a vs-bundle
on $M$, maps $f: B{\to}E$ and $f':M{\to}M$ are supersmooth, and
$(i) {\quad}{{\pi}_E}{\circ}f=f'{\circ}{{\pi}_B};$
$(ii) {\quad} f:{\pi}^{-1}_B {(x)} {\to} {\pi}^{-1}_E {\circ} f'(x)$ is a vs-space
homomorphism.
\par
We denote by $u=(x,y)=({\hat x},{\theta},{\hat y},{\zeta})=\{ u^{\alpha}=
(x^{I},y^{A})=({{\hat x}^i},{\theta}^{\hat i},{{\hat y}^a},{\zeta}^{\hat a})=
({{\hat x}^i}, x^{\hat i},{{\hat y}^a},y^{\hat a})\}$ the local coordinates
in ${\hat {\cal E}}$ and write their transformations as
\begin{equation}
x^{I'}=x^{I'}({x^I}),{\quad} srank({{\partial}x^{I'}\over{\partial}x^I})=(n,k),
\label {f1}
\end{equation}
$y^{A'}=M^{A'}_{A}(x) {y}^A,$ where $M_{A}^{A'}(x){\in} G(m,l,\Lambda).$
\par
For local coordinates and geometric objects on ts-bundle $TS$ we shall not
distinguish indices of coordinates on the base and in the fibre and write,
for instance,
$u=(x,y)=({\hat x},{\theta},{\hat y},{\zeta})=\{u^{\alpha}=({x^I},{y^I})=
({{\hat x}^i},{\theta}^{\hat i},{{\hat y}^i},{\zeta}^{\hat i})=
({{\hat x}^i},x^{\hat i},{{\hat y}^i},y^{\hat i})\}.$
\par
Finally, in this section, we remark that to simplify considerations in this
work we shall consider only locally trivial super fibre bundles.
\section{Nonlinear Connections in Vector Superbundles}

The concept of nonlinear connection (N-connection)
 was introduced in the framework of Finsler
geometry [24,41,42].The global definition  of N-connection is given in [43].
In works [26,27] nonlinear connection structures are studied in details. In
this section we shall present the notion of nonlinear connection in vs-bundles
and its main properties in a way necessary for our further considerations.
\par
Let us consider a vs-bundle ${\cal E}= (E, {\pi}_{E},M)$ whose type fibre is
${\hat{\cal F}}$ and ${{\pi}^T}:T{\cal E}{\to}TM$ is the superdifferential
of the map ${{\pi}_E}$ (${{\pi}^T}$ is a fibre-preserving morphism of the
st-bundle $(T{\cal E}, {{\tau}_E},M)$ to $E$ and of st-bundle
$(TM,{\tau},M)$ to $M$). The kernel of this vs-bundle morphism being a subbundle
of $(TE,{{\tau}_E},E)$ is called the vertical subbundle over ${\cal E}$ and denoted by
$V{\cal E}=(VE,{{\tau}_V}, E)$. Its total space is
 $V{\cal E}={{\bigcup}_{u\in{\cal E}}}
{\quad} {V_u},{\quad}$ where ${V_u}= {ker}{{\pi}^T} ,{\quad} {u{\in}{\cal E}}.$
A vector
$$ Y={Y^{\alpha}}{{\partial}\over{{\partial} {u^{\alpha}}}}=
{Y^I}{{\partial}\over{{\partial}{x^I}}}+{Y^A}{{\partial}\over{{\partial}{y^A}}}=$$
$$
{Y^i}{{\partial}\over{{\partial}{x^i}}}+{Y^{\hat i}}{{\partial}\over
{{\partial}{{\theta}^{\hat i}}}}+{Y^a}{{\partial}\over{{\partial}{y^a}}}+
{Y^{\hat a}}{\partial\over\partial {{\zeta}^{\hat a}}} $$
tangent to $\cal E$ in the point $u\in{\cal E}$ is locally represented as
$$(u,Y)=({u^{\alpha}},{Y^{\alpha}})=({x^{I}},{y^{A}},{Y^{I}},{Y^A})=$$
$$({{\hat x}^i},{{\theta}^{\hat i}},{{\hat y}^a},{{\zeta}^{\hat a}},{{\hat Y}^i},
{Y^{\hat i}},{{\hat Y}^a},{Y^{\hat a}}).$$
\begin{defin}
A nonlinear connection, N-connection, in sv-bundle ${\cal E}$ is a splitting on
the left of the exact sequence
\begin{equation}
 0{\longmapsto}{V \cal E}\stackrel{i}{\longmapsto}{T{\cal E}}
{\longmapsto}{{T{\cal E}{/}V{\cal E}}}{\longmapsto}0, \label {f2}
 \end{equation}
i.e. a morphism of vs-bundles $N: T{\cal E}\in{V{\cal E}}$ such that
$N{\circ}i$ is the identity on $V{\cal E}$.
\end{defin}
\par
The kernel of the morphism $N$ is called the horizontal subbundle and denoted by
$(HE,{{\tau}_E},E).$ From the exact sequence (2) one follows that N-connection
structure can be equivalently defined as a distribution $\{{{E_u}\to{{H_u}E}},
{{T_u}E}={{H_u}E}{\oplus}{{V_u}E}\}$ on $E$ defining a global decomposition,
as a Whitney sum,
\begin{equation}
T{\cal E}=H{\cal E}+V{\cal E}. \label {f3}
\end{equation}
\par
To a given N-connection we can associate a covariant s-derivation on M:
\begin{equation}
{\bigtriangledown}_X {Y}=X^I{\{ {{\partial Y^{A}}\over{\partial x^{I}}}+
{N^{A}_{I}}(x,Y)\}} s_A,\label {f4}
\end{equation}
where $s_A$ are local independent sections of ${\cal E},{\quad}Y={Y^A}s_A$ and
$X={X^I}s_I$.
\par
S-differentiable functions $N^{A}_{I}$ from (4) written as functions on $x^I$
and $y^{A},{\quad} N^{A}_{I}(x,y),$ are called the coefficients of the N-connection
and satisfy these transformation laws under coordinate transforms (1) in $\cal E$:
$$ N^{A'}_{I'}{{\partial x^{I'}}\over{\partial x^{I}}}=M^{A'}_{A} N^{A}_{I}-
{\partial {M^{A'}_{A}{(x)}}\over\partial x^I} {y^A}.$$
\par
If coefficients of a given N-connection are s- differentiable with respect to
coordinates $y^A$ we can introduce (additionally to covariant nonlinear
s-derivation (4)) a linear covariant s-derivation $\hat D$ (which is a generalization
for sv-bundles of the Berwald connection [44]) given as follows:
$$ {{\hat D}_{({{\partial}\over{\partial x^{I}}})}}({{\partial}\over{\partial y^{A}}})=
{{{\hat N}^{B}}_{AI}}({{\partial}\over{\partial y^{B}}}),{\quad}
{{\hat D}_{({{\partial}\over{\partial y^{A}}})}}({{\partial}\over{\partial y^{B}}})=0,$$
where
\begin{equation}
{{\hat N}^A}_{BI} (x,y) = {{{\partial}{{N^A}_I}{(x,y)}}\over{\partial y^{B}}}
\label {f5}
\end{equation}
and
$$ {{{\hat N}^A}_{BC}} {(x,y)}=0.$$
For a vector field on ${\cal E}{\quad} Z={Z^I}{\partial\over{\partial x^I}}+
{Y^A}{\partial\over{\partial y^A}}$ and $B={B^A}{(y)}{\partial
\over{\partial y^A}}$
being a section in the vertical s-bundle $(VE,{{\tau}_V},E)$ the linear connection
(5) defines s-derivation (compare with (4)):
$${{\hat D}_Z} B=[{Z^I}({\partial B^A\over\partial x^I}+{\hat N}^A_{BI} B^B ) +
Y^B {\partial B^A\over\partial y^B}] {\partial\over\partial y^A}.$$
\par
Another important characteristic of a N-connection is its curvature:
$$\Omega ={\frac{1}{2}}{\Omega}^A_{IJ} {dx^I}\land {dx^J} \otimes {\partial\over
\partial y^A}$$
with local coefficients
$${\Omega}^A_{IJ} = {\partial N^A_I\over\partial x^J} - {(-)}^{\vert IJ\vert}
{\partial N^A_J\over\partial x^I} + N^B_I {\hat N}^A_{BJ} - {(-)}^{\vert IJ\vert}
 N^B_J {\hat N}^A_{BI},$$
where for simplicity we have written ${(-)}^{{\mid K \mid}{\mid J \mid}} =
{(-)}^{\mid {KJ} \mid}.$
\par
We note that linear connections are particular cases of N-connections, when
$N^A_I {(x,y)}$ are parametrized as $N^A_I {(x,y)} = K^A_{BI} {(x)} x^I y^B,$
where functions $K^A_{BI} {(x)}$, defined on M, are called the Christoffel coefficients.

\section{Geometry of the Total Space of a Sv-Bundle}

The geometry of the sv- and st-bundles is very rich.It contains a lot of geometrical
objects and properties which could be of great importance in theoretical physics.
In this section we shall present the main results from geometry of total spaces
of sv-bundles.In order to avoid long computations and maintain the geometric
meaning the notion of nonlinear connections will systematically used in a manner
generalizing to s-spaces the classical results [26,27].
\subsection{Distinguished tensors and
connections in sv-bundles}
\label{sec:level2}
In sv-bundle $\cal E$ we can introduce a local basis adapted to the given
N-connection:
\begin{equation}
{\delta}_{\alpha} ={\delta\over\delta u^{\alpha}} = ( {{\delta}_I}
 = {\delta\over\delta x^I} =
{{\partial}_I} -{N^A_I} {(x,y)}{\partial\over\partial y^A},
{\quad} {{\partial}_A} ) , {\quad}
\label {f6}
\end{equation}
where ${\partial}_I = {\partial\over\partial x^I}$ and ${\partial}_A =
{\partial\over\partial y^A}$ are usual partial s-derivations.
The dual to (6) basis is defined as
$${\delta}^{\alpha} = {\delta u^{\alpha}} =$$
\begin{equation} ( {\delta}^I = {\delta x^I} ={dx^I},
{\delta}^A = {\delta y^A} = dy^A + {N^A_I}{(x,y)}{dx^I} ).
\label {f7}
\end{equation}
\par
By using adapted bases (6) and (7) one introduces algebra $DT({\cal E})$
of distinguished tensor s-fields (ds-fields, ds-tensors, ds-objects) on
${\cal E} , {\quad} {\cal T} = {{\cal T}^{pr}_{qs}},$ which is equivalent to the
tensor algebra of sv-bundle ${{\pi}_d}: H{\cal E}{\oplus}V{\cal E}{\to}
{\cal E},$
hereafter briefly denoted as ${{\cal E}_d}.$
An element $Q{\in}{{\cal T}^{pr}_{qs}},$ , ds-field of type
${\left(\begin{array}{cc} p & r \\ q & s \end{array}\right)},$ can be written in local form as
$$
Q ={Q^{{I_1}{\dots}{I_p}{A_1}{\dots}{A_r}}_{{J_1}{\dots}{J_q}{B_1}{\dots}{B_s}}}
{(x,y)}{{\delta}_{I_1}}\otimes {\dots} \otimes {{\delta}_{I_p}} \otimes {dx^{J_1}}
\otimes {\dots} \otimes$$
\begin{equation}
 {dx^{J_q}}\otimes {{\partial}_{A_1}} \otimes {\dots}
\otimes {{\partial}_{A_r}} \otimes {{\delta}y^{B_1}} \otimes {\dots} \otimes
{{\delta}y^{B_s}}. \label {f8}
\end{equation}
\par
In addition to ds-tensors we can introduce ds-objects with various s-group and
coordinate transforms adapted to global splitting (3).
\begin{defin}
A linear distinguished connection, d- connection, in sv- bundle $\cal E$ is a
linear connection $D$ on $\cal E$ which preserves by parallelism the horizontal
and vertical distributions in $\cal E$.
\end{defin}
\par
By a linear connection of a s-manifold we understand a linear connection in its
tangent bundle.
\par
Let denote by $\Xi (M)$ and $\Xi ({\cal E}),$ respectively, the modules of vector
fields on s-manifold $M$ and sv-bundle ${\cal E}$ and by ${\cal F}{(M)}$ and
${\cal F}{({\cal E})},$ respectively, the s-modules of functions on $M$ and
on $\cal E$.
\par
It is clear that for a given global splitting into horizontal and vertical
s-subbundles (3) we can associate operators of horizontal and vertical covariant
derivations (h- and v-derivations, denoted respectively as $D^{(h)}$ and
$D^{(v)}$) with properties:
$$ {D_X}Y = (XD)Y = {D_{hX}}Y + {D_{vX}}Y,$$
where
$$ D_X^{(h)} {Y} = D_{hX} {Y},{\quad} D_X^{(h)} f = (hX)f$$ and
$$ D_X^{(v)} {Y} = D_{vX} {Y},{\quad} D_X^{(v)} f = (vX)f,$$
for every $f\in {\cal F}(M)$ with decomposition of vectors
 $X,Y\in{\Xi}({\cal E})$
into horizontal and vertical parts,
$ X=hX+vX{\quad}$ and ${\quad} Y=hY+vY.$
\par
The local coefficients of a d- connection $ D$  in $\cal E$  with respect to the
local adapted frame (6) separate into four groups. We introduce local coefficients
$({L^I}_{JK}(u),{L^A}_{BK}(u))$ of $D^{(h)}$ such that
$$ D^{(h)}_{({\delta\over\delta x^K})} {\delta\over\delta x^J} =
{L^I}_{JK} (u) {\delta\over\delta x^I},$$
$$ D^{(h)}_{({\delta\over\delta x^K})} {\partial\over\partial y^B}=
{L^A}_{BK} {(u)} {\partial\over\partial y^A},$$
$$ D^{(h)}_{({\delta\over\delta x^k})} f ={{\delta}f\over\delta x^K}=
{\partial f\over\partial x^K} - N^A_K {(u)} {\partial f\over\partial y^A},$$
and local coefficients $({C^I}_{JC} (u), {C^A}_{BC} (u))$ such that
$$ D^{(v)}_{({\partial\over\partial y^C})} {\delta\over\delta x^J}=
{C^I}_{JC} (u) {\delta\over\delta x^I},{\quad}
D^{(v)}_{({\partial\over\partial y^C})} {\partial\over\partial y^B} =
{C^A}_{BC} {\partial\over\partial y^A},$$
$$ D^{(v)}_{({\partial\over\partial y^C})} f = {\partial f\over\partial y^C},$$
where $f\in{\cal F}({\cal E}).$
\cal
The covariant d-derivation along vector $X= X^I {\delta\over\delta x^I}+
Y^A {\partial\over\partial y^A}$ of a ds-tensor field $Q$ of type
${\left(\begin{array}{cc} p & r \\ q & s\end{array}\right)} ,{\quad}$see (8), can be written as
$${D_X} Q = D^{(h)}_X Q + D^{(v)}_X Q,$$
where h-covariant derivative is defined as
$$ D^{(h)}_X Q = X^K Q^{IA}_{JB{\mid}K} {\delta}_I {\otimes} {\partial}_A
{\otimes}dx^I{\otimes}{\delta y^A},$$
with components
\widetext
$$Q^{IA}_{JB{\mid}K} = {\delta {Q^{IA}_{JB}}
\over\delta x^K} +{L^I}_{HK} Q^{HA}_{JB}+
{L^A}_{CK} W^{IC}_{JB} - {L^H}_{JK} W^{IA}_{HB} -{L^C}_{BK} W^{IA}_{JC} ,$$
and v-covariant derivative is defined as
$$ {D^{(v)}_X} Q = {X^C} {Q^{IA}_{JB{\perp}C}} {\delta}_I {\otimes} {\partial}_A
{\otimes} dx^I {\otimes} {\delta y^B},$$
with components
$$ {Q^{IA}_{JB{\perp}C}} = {\partial {Q^{IA}_{JB}} \over\partial y^C} +
{C^I}_{HC} {Q^{HA}_{JB}} +{C^A}_{DC} {Q^{ID}_{JB}} -
{C^H}_{JC} {Q^{IA}_{HB}} -{C^D}_{BC} {Q^{IA}_{JD}}.$$
\par
The above presented formulas show that$$
D{\Gamma} = (L,{\tilde L},{\tilde C},C) =$$  $$ ({L^A}_{JK} (u), {L^A}_{BK} (u),
{C^I}_{JA} (u), {C^A}_{BC} (u))$$ are the local coefficients of the
d-connection $D$ with respect to the local frame $({\delta\over\delta x^I},
{\partial\over\partial y^a}).$ If a change (1) of local coordinates on
$\cal E$ is performed, by using the law of transformation of local frames
under it
\begin{equation}
({\quad}{\delta}_{\alpha} = ({\delta}_I , {\partial}_A) \longmapsto
{\delta}_{{\alpha}'} = ( {\delta}_{I'} ,{\partial}_{A'}),
\label {f9}
\end{equation}
where $$ {\delta}_{I'} = {\partial x^I \over\partial x^{I'}} {\delta}_I ,{\quad}
{\partial}_{A'} = M^A_{A'} (x) {\partial}_A {\quad} ),$$
we obtain the following transformation laws of the local coefficients of a
d-connection:
\begin{equation}
{L^{I'}}_{J'K'} = {\partial x^{I'} \over\partial x^I} {\partial x^J\over
\partial x^{J'}} {\partial x^K \over\partial x^{K'}} {L^I}_{JK}+
{\partial x^{I'}\over\partial x^K} {{\partial}^2 x^K \over {{\partial x^{J'}}
{\partial x^{K'}}}}, \label {f10}
\end{equation}
$$ {L^{A'}}_{B'K'}= M^{A'}_{A} M^B_{B'} {\partial x^K \over\partial x^{K'}}
{L^A}_{BK} + M^{A'}_C {\partial {M^{C}_{B'}}\over\partial x^{K'}} ,$$and$$
{C^{I'}_{J'C'}} = {\partial x^{I'} \over\partial x^I} {\partial x^J\over
\partial x^{J'}} {M^C_{C'}} {{C^I}_{JC}} ,  {C^{A'}_{B'C'}} ={M^{A'}_A}
  {M^B_{B'}}  {M^C_{C'}} {{C^A}_{BC}} .$$
As in the usual case of tensor calculus on locally isotropic spaces
the transformation laws (10) for d-connections differ from those for
ds-tensors, which are written (for instance, we consider transformation
laws for ds-tensor (8)) as
$$ Q^{{I'_1} {\dots} {A'_1} {\dots} }_{{J'_1} {\dots} {B'_1} {\dots} } =
{\partial x^{I'_1} \over \partial x^{I_1}} {\dots} M^{A'_1}_{A_1} {\dots}
{\partial x^{J_1} \over\partial x^{J'_1}} {\dots} M^{B_1}_{B'_1} {\dots}
Q^{{I_1}{\dots}{A_1}{\dots}}_{{J_1}{\dots}{B_1}{\dots}} .$$
\par
We note that defined distinguished s-tensor algebra and d-covariant calculus
in sv-bundles provided with N-connection structure is a supersymmetric
generalization of the corresponding formalism for usual vector bundles
presented in [26,27].  To obtain Miron and Anastasiei local formulas we have
to restrict us with even components of geometric objects by changing, formally,
capital indices $(I,J,K,...)$ into $(i,j,k,a,..)$ and s-derivation and s-commutation
rules into those for real number fields on usual manifolds. For brevity, in this
work we shall omit proofs and cumbersome computations if they will be simple
supersymmetric generalizations of those presented in the just cited monographs.

\subsection{Torsion and curvature of the distinguished connection in sv-bundle}

Let ${\cal E}=(E,{\pi}_E,M)$ be a sv--bundle endowed with N-connection and
d-connection structures. The torsion of d-connection is introduced into usual
manner:
$$ T(X,Y)=[X,DY\} -[X,Y \}  ,{\quad}  X,Y {\subset}  {\Xi}  {(M)}.$$
The following decomposition is possible by using h-- and  v--projections
(associated to N):
$$T(X,Y)=T(hX,hY)+T(hX,vY)+T(vX,hX)+T(vX,vY).$$
Taking into account the skewsupersymmetry of T and the equation $h[vX,vY\}=0$
we can verify that the torsion of a d-connection is completely determined by
the following ds-tensor fields:
$$hT(hX,hY)= [X (D^{(h)}{h})Y\} - h [hX,hY\} ,$$
$$vT(hX,hY)= -v[hX,hY\}, $$
$$hT(hX,vY)= - D^{(v)}_Y {hX} - h[hX,vY\}, $$
$$vT(hX,vY)= D^{(h)}_X {vY} - v[hX,vY\}, $$
$$vT(vX,xY)= [X(D^{(v)} v)Y\} - v [vX,vY\}, $$
where $X,Y\in {{\Xi}({\cal E})}.$
In order to get the local form of the ds-tensor fields which determine the
torsion of d-connection $D{\Gamma}$ (the torsions of $D{\Gamma}$) we use
equations
$$[{\delta\over\delta x^J} , {\delta\over\delta x^K} \}= {R^A}_{JK}
{\partial\over\partial y^A},$$
where
$${R^A}_{JK} ={{\delta {N^A_J}}\over\delta x^K} - {(-)}^{\mid {KJ} \mid}
{{\delta {N^A_K}}\over\delta x^J},$$
$$[{\delta\over\delta x^J}, {\partial\over\partial y^A} \} =
{\partial {N^A_J}\over\partial y^B} {\partial\over\partial y^A},$$
and introduce notations
\begin{equation}
hT({\delta\over\delta x^K},{\delta\over\delta x^J})
= {T^I}_{JK} {\delta\over\delta x^I},{\quad}
vT{({\delta\over\delta x^K},{\delta\over\delta x^J})} ={{{\tilde T}^A}_{JK}}
{\partial\over\partial y^A}, \label {f11}
\end{equation}
$$hT({\partial\over\partial y^A},{\partial\over\partial x^J})=
{{{\tilde P}^I}_{JB}} {\delta\over\delta x^I},{\quad} vT({\partial\over\partial y^B},
{\delta\over\delta x^J}) = {P^A}_{JB} {\partial\over\partial y^A},$$
$$vT({\partial\over\partial y^B},{\partial\over\partial y^B})=
{S^A}_{BC} {\partial\over\partial y^A}. $$
\par
Now we can compute the local components of the torsions, introduced in (11),
with respect to the frame $({\delta\over\delta x},{\partial\over\partial y}),$
of a d-connection $D{\Gamma}=(L,{\tilde L},{\tilde C}, C):$
\begin{equation}
{T^I}_{JK} = {L^I}_{JK} - {(-)}^{\mid {JK}\mid} {L^I}_{KJ},
{{\tilde T}^A}_{JK} = {R^A}_{JK}, {{\tilde P}^I}_{JB} ={C^I}_{JB},
\label {f12}
\end{equation}
$${P^A}_{JB} = {\partial {N^A_J}\over\partial y^B} -{L^A}_{BJ},{\quad}
{S^A}_{BC}={C^A}_{BC}-{(-)}^{\mid BC \mid} {C^A}_{CB}.$$
The even and odd components of torsions (12) can be specified in explicit
form by using decompositions of indices into even and odd parts $(I=(i,{\hat i}),
J=(j,{\hat j}),..)$, for instance,
$${T^i}_{jk} = {L^i}_{jk} -{L^i}_{kj},{\quad} {T^i}_{j{\hat k}} =
{L^i}_{j{\hat k}}+
{L^i}_{{\hat k}j},$$  $${T^{\hat i}}_{jk} = {L^{\hat i}}_{jk}-{L^{\hat i}}_{kj},
{\dots},$$ and so on.
\par
Another important characteristic of a d-connection $D\Gamma$ is its curvature:
$$R(X,Y)Z = D_{[X} D_{Y\}} - D_{[X,Y\}} Z, $$
where $X,Y,Z \in {\Xi}(E).$
By using h- and v-projections we can prove that
\begin{equation}
vR(X,Y)hZ=0, {\quad} hR(X,Y)vZ=0 \label {f13}
\end{equation}
and
$$R(X,Y)Z=hR(X,Y)hZ+vR(X,Y)vZ,$$
where $X,Y,Z \in {\Xi}(E).$
Taking into account properties (13) and the equation $R(X,Y)=-{(-)}^{\mid XY \mid}
R(Y,X)$ we prove that the curvature of a d-connection $D$ in the total space of
a sv-bundle $\cal E$ is completely determined by the following six ds-tensor
fields:
\begin{equation}
R(hX,hY)hZ = ({D^{(h)}_{[X}} D^{(h)}_{Y\}} - D^{(h)}_{[hX,hY\}} -
D^{(v)}_{[hX,hY\}})hZ, \label {f14}
\end{equation}
$$R(hX,hY)vZ = ( D^{(h)}_{[X} D^{(h)}_{Y\}} - D^{(h)}_{[hX,hY\}}
                                            - D^{(v)}_{[hX,hY\}}) vZ,$$
$$R(vX,hY)hZ = ( D^{(v)}_{[X} D^{(h)}_{Y\}} - D^{(h)}_{[vX,hY\}}
                                            - D^{(v)}_{[vX,hY\}}) hZ,$$
$$R(vX,hY)vZ = ( D^{(v)}_{[X} D^{(h)}_{Y\}} - D^{(h)}_{[vX,hY\}}
                                            - D^{(v)}_{[vX,hY\}}) vZ,$$
$$R(vX,vY)hZ = ( D^{(v)}_{[X} D^{(v)}_{Y\}} - D^{(v)}_{[vX,vY\}}) hZ,$$
$$R(vX,vY)vZ = ( D^{(v)}_{[X} D^{(v)}_{Y\}} - D^{(v)}_{[vX,vY\}}) vZ,$$
where $$  {D^{(h)}_{[X}} {D^{(h)}_{Y\}}} = {D^{(h)}_X} {D^{(h)}_Y} -
               {{(-)}^{\mid XY \mid}} {D^{(h)}_Y} {D^{(h)}_X} , $$
$$        {D^{(h)}_{[X}} {D^{(v)}_{Y\}}} = {D^{(h)}_X} {D^{(v)}_Y} -
               {{(-)}^{\mid XY \mid}} {D^{(v)}_Y} {D^{(h)}_X}  $$
 and
$$ {D^{(v)}_{[X}} {D^{(h)}_{Y\}}} = {D^{(v)}_X} {D^{(h)}_Y} -
               {(-)}^{\mid XY \mid} {D^{(h)}_Y} {D^{(v)}_X}.$$
We introduce the local components of ds-tensor fields (14) as follows:\begin{equation}
R({{\delta}_K},{{\delta}_J}) {{\delta}_H} ={{{R_H}^I}_{JK}}{{\delta}_I},{\quad}
R({{\delta}_K},{{\delta}_J}){{\partial}_B}=
      {{\tilde R}^{{\cdot}A}_{{B{\cdot}}{JK}}{{\partial}_A}},
\label {f15}
\end{equation}
$$
R({{\partial}_C},{{\delta}_K}){{\delta}_J}=
      {{\tilde P}^{{\cdot}I}_{{J{\cdot}}{KC}}{{\delta}_I},{\quad}
R({{\partial}_C},{{\delta}_K}){{\partial}_B}=
                                            {{{P_B}^A}_{KC}}}{{\partial}_A} , $$
$$
R({{\partial}_C},{{\partial}_B}){{\delta}_J}=
      {{\tilde S}^{{\cdot}I}}_{{J{\cdot}}{BC}}{{\delta}_I},{\quad}
R({{\partial}_D},{{\partial}_C}){{\partial}_B}=
                                            {{{S_B}^A}_{CD}}{{\partial}_A} . $$
Putting the components of a d-connection $D{\Gamma}=(L,{\tilde L},{\tilde C},C)$
in (15), by a direct computation, we obtain these locally adapted components
of the curvature (curvatures):
\widetext
$$
{{R_H}^I}_{JK} = {{\delta}_K} {L^I}_{HJ} - {(-)}^{\mid KJ \mid}
 {{\delta}_J} {L^I}_{HK} +
{L^M}_{HJ} {L^I}_{MK} - {(-)}^{\mid KJ \mid} {L^M}_{HK} {L^I}_{MJ} +
{C^I}_{HA} {R^A}_{JK},
$$
$$
{{\tilde R}^{\cdot A}_{B \cdot JK}} = {{\delta}_K} {L^A}_{BJ} - {(-)}^{\mid KJ \mid}
  {{\delta}_J} {L^A}_{BK} +
{L^C}_{BJ} {L^A}_{CK} - {(-)}^{\mid KJ \mid} {L^C}_{BK} {L^A}_{KJ} +
{C^A}_{BC} {R^C}_{JK} ,$$
\begin{equation}
{{\tilde P}^{\cdot I}_{J \cdot KA}} = {{\partial}_A} {L^I}_{JK} - {C^I}_{JA{\mid}K} +
{C^I}_{JB} {P^B}_{KA} , \label {f16}
\end{equation}
$${{P_B}^A}_{KC} = {{\partial}_C} {L^A}_{BK} - {C^A}_{BC{\mid}K} +
                {C^A}_{BD} {P^D}_{KC},$$
$${{\tilde S}^{\cdot I}_{J \cdot BC}} = {{\partial}_C} {C^I}_{JB} - {(-)}^{\mid BC \mid}
    {{\partial}_B} {C^I}_{JC} + {C^H}_{JB} - {(-)}^{\mid BC \mid}
               {C^H}_{JC} {C^I}_{HB} ,$$
$${{S_B}^A}_{CD} = {{\partial}_D} {C^A}_{BC} - {(-)}^{\mid CD \mid}
     {{\partial}_C} {C^A}_{BD} + {C^E}_{BC} {C^A}_{ED} - {(-)}^{\mid CD \mid}
               {C^E}_{BD} {C^A}_{EC} .$$

We can also compute even and odd components of curvatures (16) by splitting
indices into even and odd parts, for instance,
$$ {{R_h}^i}_{jk} = {{\delta}_k} {L^i}_{hj} - {{\delta}_j}{{L^i}_{hk} +
{{L^m}_{hj}} {{L^i}_{mk}} - {{L^m}_{hk}} {{L^i}_{mj}}} + {{C^i}_{ha}} {{R^a}_{jk}},$$
$${{R_h}^i}_{j{\hat k}} = {{\delta}_{\hat k}} {L^i}_{hj} +
                          {{\delta}_j} {L^i}_{h{\hat k}} +
{L^m}_{hj} {L^i}_{m{\hat k}} + {L^m}_{h{\hat k}} {L^i}_{mj} +
{C^i}_{ha} {R^a}_{j{\hat k}}{\quad} , {\dots}.$$
(we omit the formulas for the rest of even--odd components of curvatures
because we shall not use them in this work).

\subsection{Bianchi and Ricci Identities for d-Connections in SV--Bundles}

The torsion and curvature of every linear connection $D$ on sv-bundle satisfy
the following generalized Bianchi identities:
$$\sum_{SC} {[(D_X {T})(Y,Z) - R(X,Y)Z + T(T(X,Y),Z)]} =0,$$
\begin{equation} \sum_{SC} {[(D_X {R})(U,Y,Z) + R(T(X,Y)Z)U]}=0, \label {f17}
\end{equation}
where $\sum_{SC}$ means the respective supersymmretric cyclic sum over $X,Y,Z$
and $U.$ If $D$ is a d-connection, then by using (13) and
$$ v(D_X {R} ) (U,Y,hZ) = 0, {\quad} h({D_X} R(U,Y,vZ) =0,$$the identities (17) become
$$\sum_{SC}{[h({D_X}T)(Y,Z)-hR(X,Y)Z + hT(hT(X,Y),Z) + hT(vT(X,Y),Z)]}=0,$$
$$\sum_{SC}{[v({D_X}T)(Y,Z)-vR(X,Y)Z + vT(hT(X,Y),Z) + vT(vT(X,Y),Z)]}=0,$$
$$\sum_{SC}{[h({D_X}R)(U,Y,Z) + hR(hT(X,Y),Z)U + hR(vT(X,Y),Z)U]} =0,$$
\begin{equation}
\sum_{SC}{[v({D_X}R)(U,Y,Z) + vR(hT(X,Y),Z)U + vR(vT(X,Y),Z)U]} =0.\label {f18}
\end{equation}
In order to get the local adapted form of these identities we insert in (18)
these necessary values of triples $(X,Y,Z)$,(${\quad}=({{\delta}_J},{{\delta}_K},
{{\delta}_L}),$ or $({{\partial}_D},{{\partial}_C},{{\partial}_B}),$) and
putting successively $U={\delta}_H$ and $U={\partial}_A.$ Taking into account
(11),(12) and (14),(15) we obtain:
$$\sum_{SC[L,K,J\}} [{T^I}_{JK{\mid}H} + {T^M}_{JK} {T^J}_{HM} +
  {R^A}_{JK} {C^I}_{HA} - {{R_J}^I}_{KH}] =0,$$
\begin{equation}
\sum_{SC[L,K,J\}} [{{R^A}_{JK{\mid}H}} + {T^M}_{JK} {R^A}_{HM} + {R^B}_{JK}
{P^A}_{HB}] =0, \label {f19}
\end{equation}
$${C^I}_{JB{\mid}K} - {(-)}^{\mid JK \mid} {C^I}_{KB{\mid}J} - {T^I}_{JK{\mid}B}
+{C^M}_{JB} {T^I}_{KM} - {(-)}^{\mid JK \mid} {C^M}_{KB} {T^I}_{JM} +$$
$${T^M}_{JK} {C^I}_{MB} +{P^D}_{JB} {C^I}_{KD} - {(-)}^{\mid KJ \mid}
{P^D}_{KB} {C^I}_{JD} +
{{P_J}^I}_{KB} - {(-)}^{\mid KJ \mid} {{P_K}^I}_{JB} =0,$$
$$ {P^A}_{JB{\mid}K} - {(-)}^{\mid KJ \mid} {P^A}_{KB{\mid}J} - {R^A}_{JK \perp B}
+ {C^M}_{JB} {R^A}_{KM} - {(-)}^{\mid KJ \mid} {C^M}_{KB} {R^A}_{JM} +$$
$${T^M}_{JK} {P^A}_{MB} + {P^D}_{JB} {P^A}_{KD} - {(-)}^{\mid KJ \mid}
{P^D}_{KB} {P^A}_{JD} -
{{R^D}_{JK}} {{S^A}_{BD}} + {{\tilde R}^{\cdot A}_{B\cdot JK}} = 0,$$
$$ {C^I}_{JB \perp C} - {(-)}^{\mid BC \mid} {C^I}_{JC \perp B} +
{C^M}_{JC} {C^I}_{MB} - {(-)}^{\mid BC \mid} {C^M}_{JB} {C^I}_{MC} +
{S^D}_{BC} {C^I}_{JD} - {{\tilde S}^{\cdot I}_{J\cdot BC}} =0,$$
$$ {P^A}_{JB \perp C} - {(-)}^{\mid BC \mid} {P^A}_{JC \perp B} +
{S^A}_{BC \mid J} + {C^M}_{JC} {P^A}_{MB} - {(-)}^{\mid BC \mid} {C^M}_{JB} {P^A}_{MC} +$$
$${P^D}_{JB} {S^A}_{CD} - {(-)}^{\mid CB \mid} {P^D}_{JC} {S^A}_{BD} +
{S^D}_{BC} {P^A}_{JD} + {{P_B}^A}_{JC} - {(-)}^{\mid CB \mid} {{P_C}^A}_{JB} =0,$$
$$\sum_{SC[B,C,D\}} [{S^A}_{BC\perp D} + {S^F}_{BC} {S^A}_{DF} - {{S_B}^A}_{CD} ]
=0,$$
$$\sum_{SC[H,J,L\}} [{{R_K}^I}_{HJ \mid L} - {T^M}_{HJ} {{R_K}^I}_{LM} -
{{R^A}_{HJ}} {{\tilde P}^{\cdot I}_{K \cdot LA}}] =0,$$
$$\sum_{SC[H,J,L\}} [ {{\tilde R}^{\cdot A}_{D\cdot HJ \mid L}} -
{{T^M}_{HJ}} {{\tilde R}^{\cdot A}_{D\cdot LM}} - {{R^C}_{HJ}} {{{P_D}^A}_{LC}}] =0,$$
$${{\tilde P}^{\cdot I}_{K \cdot JD \mid L}} - {(-)}^{\mid LJ \mid} {{\tilde P}^{\cdot I}_{K\cdot LD \mid J}} +
{{R_K}^I}_{LJ \perp D} + {C^M}_{LD} {{R_K}^I}_{JM} -
{(-)}^{\mid LJ \mid} {C^M}_{JD} {{R_K}^I}_{LM} - $$
$$ {T^M}_{JL} {{\tilde P}^{\cdot I}_{K\cdot MD}} + {P^A}_{LD} {{\tilde P}^{\cdot I}_{K\cdot JA}} -
{(-)}^{\mid LJ \mid} {{P^A}_{JD}} {{\tilde P}^{\cdot I}_{K\cdot LA}} -
{{R^A}_{JL}} {{\tilde S}^{\cdot I}_{K\cdot AD}} =0,$$
$${{P_C}^A}_{JD \mid L} - {(-)}^{\mid LJ \mid} {{P_C}^A}_{LD \mid J} +
{{\tilde R}^{\cdot A}_{C\cdot LJ \mid D}} + {C^M}_{LD} {{R_C}^A}_{JM} - {(-)}^{\mid LJ \mid}
{C^M}_{JD} {{R_C}^A}_{LM} -$$
$${T^M}_{JL} {{P_C}^A}_{MD} + {P^F}_{LD} {{P_C}^A}_{JF} -
{(-)}^{\mid LJ \mid} {P^F}_{JD} {{P_C}^A}_{LF} - {R^F}_{JL} {{S_C}^A}_{FD} =0,$$
$${{\tilde P}^{\cdot I}_{K\cdot JD \perp C}} -
{(-)}^{\mid CD \mid} {{\tilde P}^{\cdot I}_{K\cdot JC \perp D}} + {{S_D}^I}_{DC \mid J} +
{C^M}_{JD} {{\tilde P}^{\cdot I}_{K\cdot MC}} -
{(-)}^{\mid CD \mid} {C^M}_{JC} {{\tilde P}^{\cdot I}_{K\cdot MD}} +$$
$${P^A}_{JC} {{\tilde S}^{\cdot I}_{K \cdot DA}} -
{(-)}^{\mid CD \mid} {P^A}_{JD} {{\tilde S}^{\cdot I}_{K\cdot CA}} +
 {S^A}_{CD} {{\tilde P}^{\cdot I}_{K\cdot JA}} =0,$$
$$ {{P_B}^A}_{JD \perp C} - {(-)}^{\mid CD \mid} {{P_B}^A}_{JC \perp D} +
{{S_B}^A}_{CD \mid J} + {C^M}_{JD} {{P_B}^A}_{MC} -
{(-)}^{\mid CD \mid} {C^M}_{JC} {{P_B}^A}_{MD}+$$
$${P^F}_{JC} {{S_B}^A}_{DF} - {(-)}^{\mid CD \mid} {P^F}_{JD} {{S_B}^A}_{CF} +
{S^F}_{CD} {{P_B}^A}_{JF} =0,$$
$$\sum_{SC[B,C,D\}} {[{{S_K}^I}_{BC \perp D} - {S^A}_{BC} {{\tilde S}^{\cdot I}_{K\cdot DA}} ]}=
0,$$
$$\sum_{SC[B,C,D\}} {[{{S_F}^A}_{BC \perp D} - {S^E}_{BC} {{{S_F}^A}_{DE}}]}=0,$$
where, for instanse, ${\sum_{SC[B,C,D\}}}$ means the supersymmetric cyclic sum
over indices $B,C,D.$
\par
Identities (19) can be detailed for even and odd components of d-connection,
torsion and curvature and become very simple if ${T^I}_{JK}=0$ and ${S^A}_{BC}=0,$
.\par
As a consequence of a corresponding arrangement of (14) we obtain the Ricci
identities (for simplicity we establish them only for ds-vector fields,
although they may be written for every ds-tensor field): \begin{equation}
 {D^{(h)}_{[X}} {D^{(h)}_{Y\}}} hZ = R(hX,hY)hZ +
{D^{(h)}_{[hX,hY\}}} hZ + {D^{(v)}_{[hX,hY\}}} hZ, \label {f20}
\end{equation}
$$ {D^{(v)}_{[X}} {D^{(h)}_{Y\}}} hZ = R(vX,hY)hZ +
{D^{(h)}_{[vX,hY\}}} hZ + {D^{(v)}_{[vX,hY\}}} hZ,$$
$$ {D^{(v)}_{[X}} {D^{(v)}_{Y\}}} hZ = R(vX,vY)hZ +{D^{(v)}_{[vX,vY\}}} hZ  $$
 and \begin{equation}
 {D^{(h)}_{[X}} {D^{(h)}_{Y\}}} vZ = R(hX,hY)vZ +
{D^{(h)}_{[hX,hY\}}} vZ + {D^{(v)}_{[hX,hY\}}} vZ, \label {f21}
\end{equation}
$$ {D^{(v)}_{[X}} {D^{(h)}_{Y\}}} vZ = R(vX,hY)vZ +
{D^{(v)}_{[vX,hY\}}} vZ + {D^{(v)}_{[vX,hY\}}} vZ,$$
$$ {D^{(v)}_{[X}} {D^{(v)}_{Y\}}} hZ = R(vX,vY)vZ +
{D^{(v)}_{[vX,vY\}}} vZ.$$
Considering $X={X^I}(u){\delta\over\delta x^I} + {X^A}(u){\partial\over\partial y^A}$
and taking into account the local form of the h- and v-covariant s-derivatives
and (11),(12),(14),(15) we can express respectively identities (20) and (21) in
this form:
$${X^A}_{\mid K \mid L} - {(-)}^{\mid KL \mid} {X^A}_{\mid L \mid K} =
{{{R_H}^I}_{KL}} {X^H} - {T^H}_{KL} {X^I}_{\mid H} - {R^A}_{KL} {X^I}_{\perp A},$$
$${X^I}_{\mid K \perp D} - {(-)}^{\mid KD \mid} {X^I}_{\perp D \mid K} =
{{\tilde P}^{\cdot I}_{H\cdot KD}} {X^H} - {C^H}_{KD} {X^I}_{\mid H} - {P^A}_{KD} {X^I}_{\perp A},$$
$${X^I}_{\perp B \perp C} - {(-)}^{\mid BC \mid} {X^I}_{\perp C \perp B} =
{{\tilde S}^{\cdot I}_{H\cdot BC}} {X^H} - {S^A}_{BC} {X^I}_{\perp A}$$
and
$${X^A}_{\mid K \mid L} - {(-)}^{\mid KL \mid} {X^A}_{\mid L \mid K} =
{{R_B}^S}_{KL} {X^B} - {T^H}_{KL} {X^A}_{\mid H} - {R^B}_{KL} {X^A}_{\perp B},$$
$${X^A}_{\mid K \perp B} - {(-)}^{\mid BC \mid} {X^A}_{\perp B \mid K} =
{{P_B}^A}_{KB}{X^C} - {C^H}_{KB} {X^A}_{\mid H} - {P^D}_{KB} {X^A}_{\perp B},$$
$${X^A}_{\perp B \perp C} - {(-)}^{\mid CB \mid} {X^A}_{\perp C \perp B} =
{{S_D}^A}_{BC}{X^D} - {S^D}_{BC} {X^A}_{\perp D}.$$
\subsection{Structure Equations of a d-Connection in a VS-Bundle}

Let, for instance, consider ds-tensor field:
$$ t={t^I_A} {\delta}_I {\otimes} {\delta^A}  .$$
We introduce the so-called d-connection 1-forms ${\omega}^I_J$ and
 ${{\tilde {\omega}}^A_B}$
as$$ Dt = (D{t^I_A}) {\delta}_I {\otimes} {\delta}^A $$
with $$Dt^I_A = dt^I_A + {\omega}^I_J {t^J_A} -
{{\tilde \omega }^B_A} {t^I_B} =
 t^I_{A\mid J} {dx^J} + t^I_{A\perp B} {\delta}y^B .$$
For the d-connection 1-forms of a d-connection $D$ on $\cal E$ defined by
${{\omega}^I_J}$ and ${{\tilde \omega}^A_B}$ one holds the following
structure equations:
$$ d({d^I}) - {d^H}\wedge {\omega}^I_H = -{\Omega},$$
$$ d{({{\delta}^A})} - {{\delta}^B} \wedge {{\tilde \omega}^A_B} = - {{\tilde
\Omega}^A},$$
$$ d{{\omega}^I_J} - {{\omega}^H_J}\wedge {{\omega}^I_H} = - {{\Omega}^I_J},$$
$$d{{\tilde \omega}^A_B} - {{\tilde \omega}^C_B} \wedge {{\tilde \omega}^A_C} =
- {{\tilde \Omega}^A_B},$$
in which the torsion 2-forms ${\Omega}^I$ and ${{\tilde \Omega}^A}$ are given
respectively by formulas:
$${{\Omega}^I} = {\frac{1}{2}} {T^I}_{JK} {d^J} \wedge {d^K} +
{\frac{1}{2}} {C^I}_{JK} {d^J}\wedge {{\delta}^C},$$
$${{\tilde \Omega}^A} = {\frac{1}{2}} {R^A}_{JK} {d^J}\wedge{d^K} +
{\frac{1}{2}} {P^A}_{JC} {d^J}\wedge{{\delta}^C} +
{\frac{1}{2}} {S^A}_{BC} {{\delta}^B}\wedge{{\delta}^C},$$
and
$${{\Omega}^I_J} = {\frac{1}{2}} {{R_J}^I}_{KH} {d^K}\wedge{d^H} +
{\frac{1}{2}}{{\tilde P}^{\cdot I}_{J\cdot KC}}{d^K}\wedge{{\delta}^C} +
{\frac{1}{2}}{{\tilde S}^{\cdot I}_{J\cdot KC}}{{\delta}^B}\wedge{{\delta}^C} ,$$
$${{\tilde \Omega}^A_B} = {\frac{1}{2}} {{\tilde R}^{\cdot A}_{B\cdot KH}} {d^K}\wedge{d^H} +
{\frac{1}{2}} {{P_B}^A}_{KC} {d^K}\wedge{{\delta}^C} +
{\frac{1}{2}} {{S_B}^A}_{CD} {{\delta}^C}\wedge{{\delta}^D}.$$
We have defined the exterior product on s-space  to satisfy the property \\
${{\delta}^{\alpha}} \wedge{{\delta}^{\beta}} = -{(-)}^{\mid \alpha \beta \mid}
{{\delta}^{\beta}} \wedge {{\delta}^{\alpha}}.$

\subsection{Metric Structure of the Total Spase of a SV--Bundle}

We consider the base $M$ of a vs-bundle ${\cal E} = (E, {{\pi}_E},M)$ to be
a connected and paracompact s-manifold.
\begin{defin}
A metric structure on the total space $E$ of a vs-bundle $\cal E$ is a
supersymmetric, second order, covariant s-tensor field $G$ which in every
point $u\in{\cal E}$ is given by nondegenerate s-matrix $G_{\alpha \beta} =
G({{\partial}_{\alpha}} , {{\partial}_{\alpha}} ) {\quad}$ (with nonvanishing
superdeterminant, $sdetG \not= 0 ).$
\end{defin}
\par
Similarly as for usual vector bundles [26,27] we establish this concordance
between metric and N-connection structures on ${\cal E}$:
$$ G({{\delta}_I},{{\partial}_A})=0 ,$$or,in consequence, \begin{equation}
 {G_{IA}}-{N^B_I} {h_{AB}} =0, \label {f22}
\end{equation}
where $${G_{IA}} = G ({{\partial}_I},{{\partial}_A}),$$
which gives $$ {N^B_I} = {h^{BA}} {G_{IA}},$$
where matrix $h^{AB}$ is inverse to matrix $h_{AB} =
 G({{\partial}_A},{{\partial}_B}).$
Thus, in this case, the coefficients of N-connection ${N^A_B}(u)$ are uniquely
determined by the components of the metric on $\cal E.$
\par
If the equality (22) holds, the metric on $\cal E$ decomposes as
$$ G(X,Y) = G(hX,hY) + G(vX,vY), {\quad} X,Y \in {\Xi (E)},$$and looks locally as$$
G= g_{{\alpha}{\beta}} {(u)} {{\delta}^{\alpha}}\otimes{{\delta}^{\beta}} =$$
\begin{equation}
g_{IJ}{d^I}\otimes {d^J} + h_{AB} {{\delta}^A}\otimes {{\delta}^B}. \label {f23}
\end{equation}
\begin{defin}
A d-connection $D$ on $\cal E$ is metric, or compatible with metric $G$, if
conditions $${D_{\alpha}}{G_{{\beta}{\gamma}}}=0$$
are satisfied. \end{defin} \par
We can prove that a d-connection $D$ on $\cal E$ provided with a metric $G$
is a metric d-connection if and only if
\begin{equation}
 {D^{(h)}_X} {(hG)} =0, {D^{(h)}_X} {(vG)} =0, {D^{(v)}_X} {(hG)}=0,
{D^{(v)}_X} {(vG)}=0, \label {f24}
\end{equation}
 for every $X\in {\Xi (E)}.$
Conditions (24) are written in locally adapted form as
$$g_{IJ \mid K} =0, g_{IJ \perp A} =0, h_{AB\mid K} =0, h_{AB\perp C} =0.$$\par
In the total space $E$ of sv-bundle $\cal E$ endowed with a mertic $G$ given
by (23) one exists a metric d-connection depending only on components of
G-metric and N-connection called the canonical d-connection associated to $G.$
Its local coefficients $C{\Gamma} = ({{\grave L}^I}_{JK} ,
{{\grave L}^A}_{BK} ,{{\grave C}^I}_{JC} ,{{\grave C}^A}_{BC} )$ are as follows:
\widetext
$${{\grave L}^I}_{JK} = {\frac{1}{2}} {g^{IH}}({{{\delta}_K}g_{HJ} +
{{\delta}_J}g_{HK} - {{\delta}_H}g_{JK}}),$$
\begin{equation}
{{\grave L}^A}_{BK} = {{\partial}_B}{N^A_B} + {\frac{1}{2}} {h^{AC}}
[{{{\delta}_K}{h_{BC}} - ({{\partial}_B} {N^D_K}) {h_{DC}} - ({{\partial}_C}{N^D_K})
{h_{DB}}}],  \label {f25}
 \end{equation}
$${{\grave C}^I}_{JC} = {\frac{1}{2}} {g^{IK}} {{\partial}_C}{g_{JK}},$$
$$
{{\grave C}^A}_{BC} = {\frac{1}{2}} {h^{AD}}({{{\partial}_C}{h_{DB}} +
{{\partial}_B} {h_{DC}} - {{\partial}_D}{h_{BC}}}).$$
We point out that, in general, the torsion of $C\Gamma$--connection (25) das
not vanish (see formulas (12)).
\par
It is very important to note that on sv-bundles provided with N-connection
and d-connection and metric structures realy it is defined a multiconnection
d-structure, i.e. we can use in an equivalent geometric manner different variants
of d- connections with various properties. For example, for modeling of
some physical processes we can use the Berwald type d--connection (see (5))
\begin{equation}
 B{\Gamma} = ({{L^I}_{JK}} , {{\partial}_B} {N^A_K} ,0, {{C^A}_{BC}}),
 \label {f26}
\end{equation}
where $ {L^I}_{JK} = {{\grave L}^I}_{JK}$ and $ {C^A}_{BC} = {{\grave C}^A}_{BC} ,
{\quad}  $ which is hv-metric, i.e. satisfies conditions:
$${D^{(h)}_X} hG =0 $$
and
$${D^{(v)}_X} vG =0, $$
for every $X\in \Xi {(E)},$ or in locally adapted coordinates,
$$ g_{IJ \mid K} =0 $$
and
$$ h_{AB \perp C} =0.$$
As well we can introduce the Levi-Civita connection$$\{ {{\alpha}\over {{\beta}
{\gamma}}} \} = {\frac{1}{2}} {G^{{\alpha}{\beta}} ( {{\partial}_{\beta}}
{G_{\tau \gamma}} + {{\partial}_{\gamma}} {G_{\tau \beta}} -
{{\partial}_{\tau}} {G_{\beta \gamma}} )},$$
constructed as in the Riemann geometry from components of metric $G_{{\alpha}{\beta}}$
by using partial derivations ${{\partial}_{\alpha}} = {\partial\over
\partial u^{\alpha}} = ( {\partial\over\partial x^I} , {\partial\over
\partial y^A})$ which is metric but not a d-connection.\par
Another metric d-connection can be defined as
\begin{equation}
{{{\tilde \Gamma}^{\alpha}}_{\beta \gamma}} = {\frac{1}{2}} {G^{\alpha \tau}}
({{\delta}_{\beta}} {G_{\tau \gamma}} + {{\delta}_{\gamma}}{G_{\tau \beta}} -
{{\delta}_{\tau}}{G_{\beta \gamma}} ),
  \label {f27}
 \end{equation}
with components $C{\tilde \Gamma} = ( {L^I}_{JK} , 0, 0, {C^A}_{BC} ) , $ where
coefficients ${L^I}_{JK}$ and ${C^A}_{BC}$ are computed as in formulas (26). We call
the coefficients (27) the generalized Christofell symbols on vs-bundle $\cal E.$
\par
For our further considerations it is useful to express arbitrary d-connection
as a deformation of the background d-connection (26):
 \begin{equation}
{{{\Gamma}^{\alpha}}_{\beta \gamma}} = {{\tilde \Gamma}^{\alpha}_{\cdot \beta \gamma}} +
{{P^{\alpha}}_{\beta \gamma}},
 \label {f28}
 \end{equation}
where ${{P^{\alpha}}_{\beta \gamma}}$ is called the deformation ds-tensor. Putting
splitting (29) into (12) and (16) we can express torsion
 ${T^{\alpha}}_{\beta \gamma}$
and curvature ${{R_{\beta}}^{\alpha}}_{\gamma \delta}$ of a d-connection
${{\Gamma}^{\alpha}}_{\beta \gamma}$ as respective deformations of torsion
${{\tilde T}^{\alpha}}_{\beta \gamma}$ and torsion
 ${\tilde R}^{\cdot \alpha}_{\beta \cdot \gamma \delta}$ for connection
${{\tilde \Gamma}^{\alpha}}_{\beta \gamma} {\quad}:$
\begin{equation}
{{T^{\alpha}}_{\beta \gamma}} = {{\tilde T}^{\alpha}_{\cdot \beta \gamma}} +
{{\ddot T}^{\alpha}_{\cdot \beta \gamma}}
\label {f29}
 \end{equation}
and
 \begin{equation}
{{{R_{\beta}}^{\alpha}}_{\gamma \delta}} =
{{\tilde R}^{\cdot \alpha}_{\beta \cdot \gamma \delta}} +
{{\ddot R}^{\cdot \alpha}_{\beta \cdot \gamma \delta}} ,
 \label {f30}
 \end{equation}
\widetext
where
$${{\tilde T}^{\alpha}}_{\beta \gamma} =
 {{\tilde \Gamma}^{\alpha}}_{\beta \gamma} -
{(-)}^{\mid \beta \gamma \mid} {{\tilde \Gamma}^{\alpha}}_{\gamma \beta} +
{w^{\alpha}}_{\gamma \delta} , \qquad
{{\ddot T}^{\alpha}}_{\beta \gamma} =
{{\ddot \Gamma}^{\alpha}}_{\beta \gamma} -
{(-)}^{\mid \beta \gamma \mid} {{\ddot \Gamma}^{\alpha}}_{\gamma \beta} , $$
and
$$ {{\tilde R}^{\cdot \alpha}_{\beta \cdot \gamma \delta}} = {{\delta}_{\delta}}
{{\tilde \Gamma}^{\alpha}}_{\beta \gamma} - {(-)}^{\mid \gamma \delta \mid}
{{\delta}_{\gamma}} {{\tilde \Gamma}^{\alpha}}_{\beta \delta} +
{{{\tilde \Gamma}^{\varphi}}_{\beta \gamma}} {{{\tilde \Gamma}^{\alpha}}_{\varphi
\delta}} - {(-)}^{\mid \gamma \delta \mid}
{{{\tilde \Gamma}^{\varphi}}_{\beta \delta}} {{{\tilde \Gamma}^{\alpha}}_{\varphi
\gamma}} + {{\tilde \Gamma}^{\alpha}}_{\beta \varphi} {w^{\varphi}}_{\gamma \delta},$$
$$ {{\ddot R}^{\cdot \alpha}_{\beta \cdot \gamma \delta}} =
 {{\tilde D}_{\delta}}{{P^{\alpha}}_{\beta \gamma}} -
 {(-)}^{\mid \gamma \delta \mid} {{\tilde D}_{\gamma}} {{P^{\alpha}}_{\beta \delta}} +
{{P^{\varphi}}_{\beta \gamma}} {{P^{\alpha}}_{\varphi \delta}} -
{(-)}^{\mid \gamma \delta \mid}
{{P^{\varphi}}_{\beta \delta}} {{P^{\alpha}}_{\varphi \gamma}} +
 {{P^{\alpha}}_{\beta \varphi}} {{w^{\varphi}}_{\gamma \delta}},$$
the nonholonomy coefficients ${w^{\alpha}}_{\beta \gamma}$ are defined as
$$[{\delta}_{\alpha} , {\delta}_{\beta} \} =
 {{\delta}_{\alpha}}{{\delta}_{\beta}} - {(-)}^{\alpha \beta}
 {{\delta}_{\beta}}{{\delta}_{\alpha}} =
 {w^{\tau}}_{\alpha \beta} {{\delta}_{\tau}} .$$
\par
Finally, in this section we remark that if from geometric point of view all
considered d-connections are "equal in rights" , the construction of physical
models on la-spaces requires an explicit fixing of the type of d-connection
and metric structures.

\section{Supersymmetric Extensions of Generalized Lagrange and Finsler Spaces}

Let us fix our attention to the st-bundle $TM.$The aim of this section is to
formulate some results in the supergeometry of $TM$ and to use them in order
to develop the geometry of Finsler and Lagrange superspaces (classical and
new approaches to Finsler geometry, its generalizations and applications in
physics are contained,for example, in [20-30].
\par
All presented in the previous section basic results on
sv-bundles provided with
N-connection, d-connection and metric structures hold good for $TM.$ In this
case the dimension of the base space and typical fibre coincides and we can
write locally, for instance, s-vectors as
$$ X = {X^I} {\delta}_I + {Y^I} {\partial}_I = X^I {\delta}_I + Y^{(I)} {\partial}_{(I)} ,$$
where $u^{\alpha} = ( x^I , y^J ) = ( x^I , y^{(J)} ).$
\par
On st-bundles we can define a global map
 \begin{equation}
J: \Xi (TM) \to \Xi (TM)
\label {f31}
\end{equation}
which does not depend on N-connection structure:
$$ J({\delta\over\delta x^I}) = {\partial\over\partial y^I}$$
and$$ J({\partial\over\partial y^I}) = 0.$$
This endomorphism is called the natural (or canonical) almost tangent structure
on $TM$; it has the properties:
$$ 1) J^2 =0,{\quad} 2) Im J = Ker J = VTM $$
and 3) the Nigenhuis s-tensor,
$${N_J}(X,Y) = [JX,JY \} - J[JX,Y\} - J[X,JY]$$
$$ (X,Y \in \Xi (TN)) $$
identically vanishes, i.e. the natural almost tangent structure $J$ on $TM$
is integrable.

\subsection{Notions of Generalized Lagrange, Lagrange and Finsler Superspaces}

Let $M$ be a supersmooth (n+m)-dimensional s-manifold and $(TM,{\tau},M)$ its
st-bundle. The metric of type $g_{ij} {(x,y)}$  was introduced by P.Finsler as
a generalization of that for Riemannian spaces. Variables $y=(y^i)$ can be
interpreted as parameters of local anisotropy  or of fluctuations in nonhomogeneous
and turbulent media. The most general form of metrics with local anisotropy
have been recently studied in the frame of the so-called generalized Lagrange
geometry (GL-geometry, the geometry of GL-spaces) [26,27]. For s-spaces we
introduce this
\begin{defin}
A generalized Lagrange superspace, GLS-- space, is a pair \quad  ${GL}^{n,m} =
(M, g_{IJ}(x,y))$, \quad where \quad $g_{IJ}(x,y)$ \quad is a ds-- tensor field on
\quad ${{\tilde TM} = TM - \{ 0 \} } ,$\quad supersymmetric of superrank
$(n,m).$
\end{defin}
\par
We call $g_{IJ}$ as the fundamental ds-tensor, or metric ds-tensor, of GLS-space.
In this work we shall not intrioduce a supersymmetric notion of signature in order to be
able to consider physical models with variable signature on the even part of the
s-spaces.
\par
It is well known that if $M$ is a paracompact s-manifold there exists at
least a nonlinear connection in the its tangent bundle. Thus it is quite
natural to fix a nonlinear connection N in TM and to relate it to $g_{IJ}{(x,y)},$
by using equations (22) written on TM. For simplicity, we can consider N-connection
with vanishing torsion , when $$ {\partial}_K {N^I_J} -
{(-)}^{\mid JK \mid} {\partial}_J {N^I_K} =0.$$
Let denote a normal d-connection, defined by using N and adapted to the almost
tangent structure (31) as $D{\Gamma} = ( {L^A}_{JK} , {C^A}_{JK} ) .$ This d-connection
is compatible with metric $g_{IJ} {(x,y)}$ if
$g_{IJ\mid K} =0$ and $g_{IJ\perp K}=0.$
\par
There exists an unique d-connection $C \Gamma (N)$ which is compatible with
$g_{IJ} {(u)}$ and has vanishing torsions ${T^I}_{JK}$ and ${S^I}_{JK}$ (see
formulas (12) rewritten for st-bundles). This connection, depending only on
$g_{IJ} {(u)}$ and ${N^I_J}{(u)}$ is called the canonical metric d-connection of
GLS-space. It has coefficients
\begin{equation}
{L^I}_{JK} = {\frac{1}{2}} {g^{IH}} ( {\delta}_J {g_{HK}} + {\delta}_H {g_{JK}} -
{\delta}_H {g_{JK}}), \label {f32}
\end{equation}
$${C^I}_{JK} = {\frac{1}{2}} {g^{IH}} ({\partial}_J {g_{HK}} +
{\partial}_H {g_{JK}} - {\partial}_H {g_{JK}}).$$
Of course, metric d-connections different from $C\Gamma (N)$ may be found. For
instance, there is a unique normal d-connection $D\Gamma (N) =
 ( {\bar L}^I_{\cdot JK} , {\bar C}^I_{\cdot JK} )$ which is metric and has
a priori given torsions ${T^I}_{JK}$ and ${S^I}_{JK}.$ The coefficients of
$D\Gamma (N)$ are the following ones:
$${\bar L}^I_{\cdot JK}={L^I}_{JK} - \frac{1}{2} g^{IH} ( g_{JR} {T^R}_{HK} +
g_{KR} {T^R}_{HJ} - g_{HR} {T^R}_{KJ} ),$$
$${\bar C}^I_{\cdot JK}={C^I}_{JK} - \frac{1}{2} g^{IH} ( g_{JR} {S^R}_{HK} +
g_{KR} {S^R}_{HJ} - g_{HR} {S^R}_{KJ} ),$$
where ${L^I}_{JK}$ and ${C^I}_{JK}$ are the same as for the $C\Gamma (N)$--connection
(32). \par
The Lagrange spaces were introduced [46] in order to geometrize the concept of
Lagrangian in mechanics. The Lagrange geometry is studied in details in [26,27].
For s-spaces we present this generalization:
\begin{defin}
A Lagrange s-space, LS-space, $L^{n,m} = (M, g_{IJ} ),$ is defined as a particular
case of GLS-space when the ds-metric on M can be expressed as
\begin{equation}
g_{IJ} {(u)} = {\frac{1}{2}} {{\partial}^2 {\cal L}\over
{{\partial y^I} {\partial y^J}}} , \label {f33}
\end{equation}
where ${\cal L}: TM \to \Lambda ,$ is a s-differentiable function called a s-Lagrangian
on M.
\end{defin}
\par
Now we consider the supersymmetric extension of the Finsler space:
\par
A Finsler s-metric on M is a function $ F_S : TM \to \Lambda $ having the
properties:
\par
1. The restriction of $F_S$ to $ {\tilde {TM}} = TM\setminus \{ 0 \}$ is of the
class $G^{\infty}$ and F is only supersmooth on the image of the null cross--section
in the st-bundle to M.
\par
2. The restriction of F to ${\tilde {TM}}$ is positively homogeneous of degree 1
with respect to ${(y^I)}$, i.e. $F(x,{\lambda}y) = {\lambda}F(x,y),$ where
${\lambda}$ is a real positive number.
\par
3. The restriction of F to the even subspace of $\tilde {TM}$ is a positive
function.
\par
4. The quadratic form on ${\Lambda}^{n,m}$ with the coefficients
\begin{equation}
g_{IJ} {(u)} = {\frac{1}{2}}{{\partial}^2 F^2 \over {{\partial y^I} {\partial y^J}}}
\label {f34}
\end{equation}
defined on $\tilde {TM}$ is nondegenerate.
\begin{defin}
A pair $F^{n,m} = (M,F)$ which consists from a supersmooth s-manifold M and a
Finsler s-metric is called a Finsler superspace, FS-space.
\end{defin}
\par
It's obvious that FS-spaces form a particular class of LS-spaces with s-Lagrangian
${\cal L} = {F^2}$ and a particular class of GLS-spaces with metrics of type (34).
\par
For a FS-space we can introduce the supersymmetric variant of nonlinear
Cartan connection [24,25]:
$$ N^I_J {(x,y)} = {\partial\over\partial y^J} G^{\ast I} ,$$
where $$ G^{\ast I} = {\frac{1}{4}} g^{\ast IJ} ( {{\partial}^2 {\varepsilon}
\over {\partial y^I}{\partial x^K}} {y^K} - {\partial {\varepsilon}\over
\partial x^J} ), {\quad} {\varepsilon} {(u)} = g_{IJ} {(u)} y^I y^J ,$$
and $g^{\ast IJ}$ is inverse to $ g^{\ast}_{IJ} {(u)} = {\frac{1}{2}}
{{\partial}^2 \varepsilon\over{{\partial y^I}{\partial y^J}}}.$ In this case
the coefficients of canonical metric d-connection (32) gives the supersymmetric
variants of coefficients of the Cartan connection of Finsler spaces. A similar
remark applies to the Lagrange superspaces.

\subsection{The Supersymmetric Almost Hermitian Model of the GLS--Space}

Consider a GLS--space endowed with the canonical metric d-connection $C\Gamma (N) .$
Let ${\delta}_{\alpha} = ( {\delta}_{\alpha} , {\dot \partial}_I )$ be a usual
adapted frame (6) on TM and ${\delta}^{\alpha} = ( {\partial}^I ,
 {\dot \delta}^I )$ its dual. The linear operator
$$ F: \Xi ({\tilde {TM}})\to \Xi ({\tilde {TM}}), $$
acting on ${\delta}_{\alpha}$ by $F({\delta}_I = - {\dot \partial}_I ,
F({\dot \partial}_I ) = {\delta}_I ,$ defines an almost complex structure on
${\dot {TM}}.$ We shall obtain a complex structure if and only if the even
component of the horizontal distribution N is integrable. For s-spaces, in
general with even and odd components, we write the supersymmetric almost Hermitian
property (almost Hermitian s-structure) as
$$ {F^{\alpha}_{\beta}} {F^{\beta}_{\delta}} = - {(-)}^{\mid \alpha \delta \mid}
{{\delta}^{\alpha}_{\beta}}.$$
\par
The s-metric $g_{IJ} {(x,y)}$ on GLS-spaces induces on $\dot {TM}$ the following
metric:
\begin{equation}
G= g_{IJ} {(u)} dx^I \otimes dx^J + g_{IJ} {(u)} {\delta y^I} \otimes {\delta y^J} .
\label {f35}
\end{equation}
We can verify that pair $(G,F)$ is an almost Hermitian s-structure on ${\dot {TM}}$
with the associated supersymmetric 2-form
$$\theta = g_{IJ} {(x,y)} {\delta y^I} \wedge dx^J .$$
\par
The almost Hermitian s-space $H^{2n,2m}_S = (TM,G,F),$
 provided with a metric of type (35) is called
the lift on TM, or the almost Hermitian s-model, of GLS-space $GL^{n,m}.$ We say
that a linear connection $D$ on ${\dot {TM}}$ is almost Hermitian supersymmetric of
Lagrange type if it preserves by parallelism the vertical distribution V and is
compatible with the almost Hermitian s-structure $(G,F)$, i.e.
\begin{equation}
D_X G =0, \quad  D_X F =0, \label {f36}
\end{equation}
for every $X\in \Xi (TM).$
\par
There exists an unique almost Hermitian connection of Lagrange type $D^{(c)}$
having h(hh)- and v(vv)--torsions equal to zero. We can prove (similarly as in
[26,27]) that coefficients ${( {L^I}_{JK} , {C^I}_{JK} )}$ of $D^{(c)}$ in the
adapted basis $( {\delta}_I , {\dot \delta}_J )$ are just the coefficients (32)
of the canonical metric d-connection $C \Gamma (N)$ of the GLS-space $GL^{(n,m)} .$
Inversely , we can say that $C \Gamma (N)$--connection determines on ${\tilde {TN}}$
and supersymmetric almost Hermitian connection of Lagrange type with vanishing
h(hh)- and v(vv)-torsions.
If instead of GLs-space metric $g_{IJ}$ in (34) the Lagrange (or Finsler)
s-metric  (32) (or (33)) is taken, we obtain the almost Hermitian s-model of
Lagrange (or Finsler) s-spaces $L^{n,m}$ (or ${F^{n,m}}).$
\par
We note that the natural compatibility conditions (36) for the metric (35) and
$C\Gamma (N)$--connections on $H^{2n,2m}$--spaces plays an important role for
developing physical models on la--superspaces. In the case of usual locally
anisotropic spaces geometric constructions and d--covariant calculus are
very similar to those for the Riemann and Einstein--Cartan spaces. This is
exploited for formulation in a selfconsistent manner the theory of spinors on
la-spaces [35], for introducing a geometric background for locally anisotropic
Yang--Mills and gauge like gravitational interactions [31,32] and for extending
the theory of stochastic processes and diffusion to the case of locally anisotropic
spaces and interactions on such spaces [47]. In a similar manner we
 shall use in this work N--lifts to sv- and
st-bundles in order to investigate supergravitational la--models.

\section{SUPERGRAVITY ON LOCALLY ANISOTROPIC SUPERSPACES}

In this section we shall introduce a set of Einstein and (equivalent in our
 case) gauge like gravitational equations, i.e. we shall formulate two variants
of la--supergravity, on the total
space E of a sv-bundle $\cal E$ over a supersmooth manifold M. The first model will be
a variant of locally anisotropic supergravity theory generalizing the Miron
and Anastasiei model [26,27] on vector bundles (they considered prescribed components of
N-connection and h(hh)-  and v(vv)--torsions, in our approach we shall introduce
algebraic equations for torsion and its source). The second model will be a
la--supersymmetric extension of constructions for gauge la-gravity [31,32] and
affine--gauge interpretation of the Einstein gravity [55,56]. There are two
 ways in developing supergravitational models. We can try to maintain similarity
to Einstein's general relativity (see in [48,49] an example of locally isotropic
supergravity) and  to formulate  a variant of Einstein--Cartan theory on
 sv--bundles, this will be the aim of the subsection A,  or to introduce into
 consideration supervielbein variables and to formulate a supersymmetric gauge
like model of la-supergravity (this approach is more accepted in the usual
locally isotropic supergravity, see as a review [45]). The last  variant will
 be analysed in subsection B  by using the s-bundle of supersymmetric affine
 adapted frames on la-superspaces. For both models of la--supergravity we shall
consider  the matter field contributions as giving rise to corresponding sources
in la-supergravitational field equations. A detailed study of supersymmetric
of la--gravitational and matter fields is a matter of our further investigations [58].

\subsection{Einstein--Cartan Equations on SV--Bundles}

Let consider a sv--bundle ${\cal E} = (E,{\pi}, M)$ provided with some compatible
nonlinear connection N, d--connection D    and metric G structures.For a locally
N-adapted frame we write
$$D_{({\delta\over\delta u^{\gamma}})} {\delta\over\delta u^{\beta}} =
{{\Gamma}^{\alpha}_{\beta \gamma}} {\delta\over\delta u^{\alpha}},$$
where the d-connection $D$ has the following coefficients:
\widetext
$${{{\Gamma}^I}_{JK}} =
 {{L^I}_{JK}} ,$$
\begin{equation}
 {{{\Gamma}^I}_{JA}} = {{C^I}_{JA}} ,
{{\Gamma}^I}_{AJ} = 0 , {{\Gamma}^I}_{AB} = 0,
{{\Gamma}^A}_{JK} = 0 , {{\Gamma}^A}_{JB} = 0, {{\Gamma}^A}_{BK} = {L^A}_{BK} ,
{{{\Gamma}^A}_{BC}} = {{C^A}_{BC}} . \label {f37}
\end{equation}
The nonholonomy coefficients ${{w^{\gamma}}_{\alpha \beta}},$
 defined as $[ {\delta}_{\alpha} , {\delta}_{\beta} \} =
 {{w^{\gamma}}_{\alpha \beta}} {\delta}_{\gamma} ,$ are as follows:
$${{w^K}_{IJ}} =0 , {{w^K}_{AJ}} =0, {{w^K}_{IA}} =0, {{w^K}_{AB}} =0,
{{w^A}_{IJ}} = {R^A}_{IJ} ,$$
$$ {{w^B}_{AI}} = - (-)^{|IA|} {\partial {N^B_A}\over\partial y^A} ,
 {{w^B}_{IA}} = {\partial {N^B_A}\over\partial y^A} , {{w^C}_{AB}} =0.$$
By straightforward calculations we can obtain respectively these components of
torsion, ${\cal T} ({\delta}_{\gamma} , {\delta}_{\beta} ) =
 {{\cal T}^{\alpha}_{\cdot \beta \gamma}} {\delta}_{\alpha},$
and curvature, ${\cal R} ({\delta}_{\beta} , {\delta}_{\gamma} ) {\delta}_{\tau} =
{{\cal R}^{\cdot \alpha}_{\beta \cdot \gamma \tau}} {{\delta}_{\alpha}}, $ ds-tensors:
\begin{equation}
{\cal T}^I_{\cdot JK} = {T^I}_{JK} , {\cal T}^I_{\cdot JA} = {C^I}_{JA} ,
{\cal T}^I_{\cdot JA} = - {C^I}_{JA} , {\cal T}^I_{\cdot AB} =0, \label {f38}
\end{equation}
$$ {\cal T}^A_{\cdot IJ} = {R^A}_{IJ} , {\cal T}^A_{\cdot IB} = - {P^A}_{BI} ,
 {\cal T}^A_{\cdot BI} = {P^A}_{BI} , {\cal T}^A_{\cdot BC} = {S^A}_{BC} $$ and
\begin{equation}
{\cal R}^{\cdot J}_{I \cdot KL} = {{R_J}^I}_{KL} , {\cal R}^{\cdot J}_{B \cdot KL} = 0,
 {\cal R}^{\cdot A}_{J \cdot KL} = 0 ,
{\cal R}^{\cdot A}_{B \cdot KL} = {\tilde R}^{\cdot A}_{B \cdot KL} , \label {f39}
\end{equation}
$$ {\cal R}^{\cdot I}_{J \cdot KD} = {{P_J}^I}_{KD} , {\cal R}^{\cdot A}_{B \cdot KD} = 0 ,
{\cal R}^{\cdot A}_{J \cdot KD} = 0 , {\cal R}^{\cdot A}_{B \cdot KD} = {{P_B}^A}_{KD} ,$$
$${\cal R}^{\cdot I}_{J \cdot DK} = - {{P_J}^I}_{KD} , {\cal R}^{\cdot I}_{B \cdot DK} = 0 ,
{\cal R}^{\cdot A}_{J \cdot DK} =0, {\cal R}^{\cdot H}_{B \cdot DK} =-{{P_B}^A}_{KD} ,$$
$$ {\cal R}^{\cdot I}_{J \cdot CD} = {{S_J}^I}_{CD} , {\cal R}^{\cdot I}_{B \cdot CD} =0,
{\cal R}^{\cdot A}_{J \cdot CD} = 0, {\cal R}^{\cdot A}_{B \cdot CD} = {{S_B}^A}_{CD} $$
(for explicit dependencies of components of torsions and curvatures on  components
of d-connection see formulas (12) and (16)).
\par
The locally adapted components ${\cal R}_{\alpha \beta} = {\cal R}ic (D)
( {\delta}_{\alpha} , {\delta}_{\beta} ) $ (we point that in general on st-bundles
${\cal R}_{\alpha \beta} \ne {(-)}^{\mid \alpha \beta \mid} {\cal R}_{\beta \alpha} )$
of the Ricci tensor are as follows:
\begin{equation}
{\cal R}_{IJ} = {{R_I}^K}_{JK} , {\cal R}_{IA} = - {}^{(2)}{P_{IA}} =
 - {\tilde P}^{\cdot K}_{I \cdot KA}  \label {f40}
\end{equation}
$${\cal R}_{AI} = {}^{(1)}{P_{AI}} = {\tilde P}^{\cdot B}_{A \cdot IB} ,
{\cal R}_{AB} = {{S_A}^C}_{BC} = S_{AB} .$$
\cal
For scalar curvature, $ {\check {\cal R}} = Sc(D) = G^{\alpha \beta}
 {\cal R}_{\alpha \beta},$ we have
\begin{equation}
Sc(D) = R + S , \label {f41}
\end{equation}
where $R = g^{IJ} R_{IJ}$ and $ S = h^{AB} S_{AB}.$
\par
The Einstein--Cartan equations on sv-bundles are written as
\begin{equation}
{\cal R}_{\alpha \beta} - {\frac{1}{2}} G_{\alpha \beta} {\check {\cal R}} +
{\lambda} G_{\alpha \beta} = {\kappa}_1 {\cal J}_{\alpha \beta} , \label {f42}
\end{equation}
and
\begin{equation}
T^{\alpha}_{\cdot \beta \gamma} + {{G_{\beta}}^{\alpha}} {{T^{\tau}}_{\gamma \tau}} -
 {(-)}^{\mid \beta \gamma \mid} {{G_{\gamma}}^{\alpha}} {{T^{\tau}}_{\beta \tau}} =
{\kappa}_2 {Q^{\alpha}}_{\beta \gamma} , \label {f43}
\end{equation}
where ${\cal J}_{\alpha \beta}$ and ${Q^{\alpha}_{\beta \gamma}}$ are respectively
components of energy-momentum and spin-density of matter ds--tensors on la-space,
${\kappa}_1$ and ${\kappa}_2$ are the corresponding interaction constants and
${\lambda}$ is the cosmological constant.
To write in a explicit form the mentioned matter sources of la-supergravity in (42)
and (43) there are necessary more detailed studies of models of interaction of
superfields on la--superspaces (see first results for Yang--Mills and spinor
fields on la-spaces in [31,32,35] and, from different points of view, [28,29,38]).
We omit such considerations in this paper.
\par
Equations (42), specified in (x,y)--components,
\begin{equation}
R_{IJ} - {\frac{1}{2}} (R+S-{\lambda}) g_{IJ} = {\kappa}_1 {\cal J}_{IJ} ,
 {}^{(1)}P_{AI} = {\kappa}_1 {}^{(1)}{\cal J}_{AI} , \label {f44}
\end{equation}
$$S_{AB} - {\frac{1}{2}} (S + R - {\lambda}) h_{AB} = {{\kappa}_2}
 {\tilde {\cal J}}_{AB} ,
 {}^{(2)}P_{IA} = - {{\kappa}_2} {}^{(2)}{{\cal J}_{IA}} , $$
are a supersymmetric, with cosmological term, generalization of the similar ones
presented in [26,27], with prescribed N-connection and h(hh)-- and v(vv)--torsions.
We have added algebraic equations (43) in order to close the system of s--gravitational
field equations (really we have also to take into account the system of
constraints (22) if locally anisotropic s--gravitational field
is associated to a ds-metric (23)).
\par
We point out that on la--superspaces the divergence $D_{\alpha} {\cal J}^{\alpha}$
 does not vanish (this is a consequence of generalized Bianchi and Ricci
identities (17),(19) and (20),(21)). The d-covariant derivations of the left
and right parts of (42), equivalently of (44), are as follows:
\widetext
$$D_{\alpha} [ {\cal R}_{\beta}^{\cdot \alpha} - {\frac{1}{2}} ( {\check {\cal R}} -
 2{\lambda} ) {\delta}_{\beta}^{\cdot \alpha} ] =
\left\{ \begin{array}{rl}
{[ {{R_J}^I} - {\frac{1}{2}} {({R + S - 2{\lambda})}}{{{\delta}_J}^I}]}_{\mid I} +
 {}^{(1)}{P^A}_{I \perp A } = 0,  \\
{[ {{S_B}^A} - {\frac{1}{2}}{({R + S - 2{\lambda})}}{{{\delta}_B}^A}]}_{\perp A} -
 {}^{(2)}{P^I}_{B \mid I} =0,
\end{array} \right.
$$
where $$ {}^{(1)}{P^A}_J = {}^{(1)}{P_{BJ}} h^{AB}, \quad
         {}^{(2)}{P^I}_B = {}^{(2)}{P_{JB}} g^{IJ}, \quad
       {R^I}_J = {R_{KJ}} g^{IK} , \quad {S^A}_B = {S_{CB}} h^{AC} ,$$
and
\begin{equation}
{D_{\alpha}} {\cal J}^{\alpha}_{\cdot \beta} = {\cal U}_{\alpha} , \label {f45}
\end{equation}
where $$ D_{\alpha} {\cal J}^{\alpha}_{\cdot \beta} =
\left\{ \begin{array}{rl}
{{\cal J}^I_{\cdot J \mid I} + {}^{(1)}{\cal J}^A_{\cdot J \perp A}} =
 {\frac{1}{{\kappa}_1}}{{\cal U}_I},\\
{{}^{(2)}{\cal J}^I_{\cdot A\mid I} +  {\cal J}^B_{\cdot A \perp B}} =
 {\frac{1}{{\kappa}_1}}{{\cal U}_A} ,
\end{array} \right. $$
and
\begin{equation}
{\cal U}_{\alpha} = {\frac{1}{2}}( G^{\beta \delta}
{{\cal R}_{\delta \cdot \varphi \beta}^{\cdot \gamma}}
{{\cal T}^{\varphi}_{\cdot \alpha \gamma}} - {(-)}^{\mid \alpha \beta \mid} G^{\beta \delta}
{{\cal R}_{\delta \cdot \varphi \alpha}^{\cdot \gamma}}
{{\cal T}^{\varphi}_{\cdot \beta \gamma}} + {{\cal R}^{\beta}_{\cdot \varphi}}
{{\cal T}^{\varphi}_{\cdot \beta \alpha}} ) . \label {f46}
\end{equation}
>From the last formula it follows that ds-vector ${\cal U}_{\alpha}$ vanishes if
d-connection (37) is torsionless.
\par
No wonder that conservation laws for values of energy--momentum type, being a
consequence of global automorphisms of spaces and s--spaces, or, respectively,
of theirs tangent spaces and s--spaces (for models on curved spaces and
s--spaces), on la--superspaces are more sophisticated because, in general, such
automorphisms do not exist for a generic local anisotropy. We can construct a
la--model of supergravity, in a way similar to that for the Einstein theory if
instead an arbitrary metric d--connection the generalized Christoffel symbols
${\tilde {\Gamma}}^{\alpha}_{\cdot \beta \gamma}$ (see (27)) are used. This will
 be a locally anisotropic  supersymmetric model on the base s-manifold M which
looks like locally isotropic on the total space  of a sv-bundle. More general
supergravitational models which are locally anisotropic on the both base and
total spaces can be generated by  using deformations of d-connections of type
(28). In this case the vector ${\cal U}_{\alpha}$ from (46) can be interpreted
as a corresponding source of generic local anisotropy satisfying generalized
conservation laws of type (45).
\par
More completely the problem of formulation of conservation laws  for
 both locally isotropic and anisotropic supergravity can be solved in the frame of
 the theory of nearly autoparallel maps of sv-bundles (with specific deformation
of d-connections (28), torsion (29) and curvature (30)), which have to generalize
our constructions from [33,34,51]. This is a matter of our further investigations.
\par
 We end this subsection with the remark that  field equations
 of type (42), equivalently (44), and (43) for la-supergravity can be similarly
introduced for the particular cases of GLS--spaces with metric (35) on
$\tilde {TM}$ with coefficients parametrized as for the Lagrange, (33), or
Finsler, (34), spaces.

\subsection{Gauge Like Locally Anisotropic Supergravity}

The great part of theories of locally isotropic s-gravity are formulated as
gauge supersymmetric models based on supervielbein formalism (see [45,51--53]). Similar approaches to la-supergravity on vs-bundles can be
developed by considering an arbitrary adapted to N-connection frame
$B_{\underline \alpha} (u) = ( B_{\underline I} (u) , B_{\underline C} (u) ) $
on $\cal E$ and supervielbein, s-vielbein, matrix
$$ {A_{\alpha}}^{\underline \alpha} (u) = {\left( \begin{array}{cc}
{A_I}^{\underline I} (u) & 0 \\ 0 & {A_C}^{\underline C} \end{array} \right ) }
\subset GL^{m,l}_{n,k}(\Lambda ) =$$
$$ GL (n,k, {\Lambda} ) \oplus GL (m,l, {\Lambda} ) $$
for which
$${\delta \over\delta u^{\alpha}} = {A_{\alpha}}^{\underline \alpha} (u)
B_{\underline \alpha} (u) ,$$
or, equivalently, $ {\delta\over\delta x^I} = {A_I}^{\underline I} (x,y)
B_{\underline I} (x,y)$ and $ {\partial\over\partial y^C} = {A_C}^{\underline C}
 B_{\underline C} (x,y) ,$ and
$$ G_{\alpha \beta} (u) = {A_{\alpha}}^{\underline \alpha} (u)
 {A_{\beta}}^{\underline \beta} (u) {\eta}_{{\underline \alpha}{\underline \beta}} ,$$
where, for simplicity, ${\eta}_{\underline \alpha \underline \beta}$ is a constant
metric on vs-space $V^{n,k} \oplus V^{l,m} .$
\par
We denote by $LN( \cal E ) $ the set of all adapted frames in all points of
sv-bundle $\cal E .$ Considering the surjective s-map ${\pi}_L$ from $LN( \cal E )$
to $\cal E$  and treating $GL^{m,l}_{n,k} ( \Lambda ) $ as the structural s-group
we define a principal s--bundle, $${\cal L}N ({\cal E}) = ( LN( {\cal E} ),
{\pi}_L : LN ( {\cal E} ) \to {\cal E} , GL^{m,l}_{n,k} ( {\Lambda} ) ),$$
called as the s--bundle  of linear adapted frames on $\cal E .$
\par
 Let denote the canonical basis of the sl-algebra ${\cal G}^{m,l}_{n,k}$ for a
s-group $GL^{m,l}_{n,k} ( {\Lambda} ) $ as $I_{\hat \alpha} ,$ where index
${\hat \alpha} = ( {\hat I}, {\hat J} )$ enumerates the $Z_2$ --graded components.
The structural coefficients ${f_{\hat \alpha \hat \beta}}^{\hat \gamma} $ of
${\cal G}^{m,l}_{n,k}$ satisfy  s-commutation rules
$$[ I_{\hat \alpha} , I_{\hat \beta} \} = {f_{\hat \alpha \hat \beta}}^{\hat \gamma}
I_{\hat \gamma} .$$
On $\cal E$ we consider the connection 1--form
\begin{equation}
{\Gamma} = {{\Gamma}^{\underline \alpha}}_{{\underline \beta} \gamma}
 (u) I^{\underline \beta}_{\underline \alpha}  du^{\gamma} , \label {f47}
\end{equation}
where $$ {{\Gamma}^{\underline \alpha}}_{\underline \beta \gamma} (u) =
{A^{\underline \alpha}}_{\alpha} {A^{\beta}}_{\underline \beta} {{\Gamma}^{\alpha}}_
{\beta \gamma} + {A^{\underline \alpha}} {\delta\over\delta u^{\gamma}} {A^{\alpha}}_
{\underline \beta} (u) , $$
${{\Gamma}^{\alpha}}_{\beta \gamma}{\quad} $
are the components of the metric d--connection (37), s-matrix
 ${\quad}{A^{\beta}}_{\underline \beta}{\quad}$
is inverse to the s-vielbein matrix $ {\quad} {A^{\underline \beta}}_{\beta} , \quad$
and $ {\quad}I^{\underline \alpha}_{\underline \beta} =
\left ( \begin{array}{cc} I^{\underline I}_{\underline J} & 0 \\ 0 &
I^{\underline A}_{\underline B} \end{array} \right ) $ is the standard
distinguished basis in SL--algebra ${\cal G}^{m,l}_{n,k} .$
\par
The curvature $\cal B$ of the connection (47),
\begin{equation}
{\cal B} = d{\Gamma} + {\Gamma} \land {\Gamma} =
{\cal R}^{\cdot \underline \beta}_{\underline \alpha \cdot \gamma \delta}
I^{\underline \alpha}_{\underline \beta}
 {\delta u^{\gamma}} \land {\delta u^{\delta}} \label {f48}
\end{equation}
has coefficients $$ {\cal R}^{\cdot \underline \beta}_{\underline \alpha \cdot
\gamma \delta} = {A^{\alpha}}_{\underline \alpha} (u) {A^{\underline \beta}}_{\beta} (u)
{\cal R}^{\cdot \beta}_{\alpha \cdot \gamma \delta} , $$ where
${\cal R}^{\cdot \beta}_{\alpha \cdot \gamma \delta}$ are the components of the
ds--tensor (39).
\par
Aside from ${\cal L}N ({\cal E})$ with vs--bundle ${\cal E}$ it is naturally
related another s--bundle, the bundle of adapted affine frames ${\cal E} N ( {\cal E} ) =
 ( AN ( {\cal E} ) , {{\pi}_A} : AN( {\cal E} ) \to {\cal E} ,
 {{AF}^{m,l}_{n,k}} ( {\Lambda} ))$
with the structural s--group ${AN}^{m,l}_{n,k} ( \Lambda ) = GL^{m,l}_{n,k} ( \Lambda )
\odot {\quad} {\Lambda}^{n,k} \oplus {\Lambda}^{m,l} $ being a semidirect
product (denoted by $ \odot$ ) of $GL^{m,l}_{n,k} ( \Lambda ) $ and
${\Lambda}^{n,k} \oplus {\Lambda}^{m,l} .$
Because as a linear s-space the LS--algebra ${{\cal A}f}^{m,l}_{n,k}$ of s--group
$AF^{m,l}_{n,k} ({\Lambda}) ,$ is a direct sum of ${\cal G}^{m,l}_{n,k}$ and
${\Lambda}^{n,k}\oplus{\Lambda}^{m,l}$ we can write forms on ${\cal A}N({\cal E})$
as $\Theta = ({\Theta}_1 , {\Theta}_2 ) ,$ where ${\Theta}_1$ is the
${\cal G}^{m,l}_{n,k}$--component and ${\Theta}_2$ is the $({\Lambda}^{n,k} \oplus
{\Lambda}^{m,l})$--component of the form $\Theta .$
The connection (47) in ${\cal L}N({\cal E})$ induces a Cartan connection
$\overline {\Gamma}$ in ${\cal A}N({\cal E})$
(see, for instance, in [55] the case of
usual affine frame bundles ). This is the unique connection on s--bundle
 ${\cal A}N({\cal E})$ represented as $ i^{\ast} {\overline \Gamma} = ( {\Gamma},
{\chi}) ,$ where $\chi$ is the shifting form and $ i : {\cal A}N({\cal E}) \to
{\cal L}N({\cal E})$ is the trivial reduction of s--bundles. If $B = ( B_{\underline
\alpha} )$ is a local adapted frame in ${\cal L}N({\cal E})$ then
${\overline B} = i \circ B $ is a local section in  ${\cal A}N ({\cal E})$ and
\begin{equation}
{\overline \Gamma} = B \Gamma = ( \Gamma , \chi ) , \label {f49}
\end{equation}
$${\overline {\cal B}} = {\overline B} {\cal B} = ( {\cal B} ,{\cal T} ), $$
where ${\chi} = e_{\underline \alpha} \otimes  {A^{\underline \alpha}}_{\alpha}
du^{\alpha} , {\quad} e_{\underline \alpha} $ is the standard basis in
${\Lambda}^{n,k} \oplus {\Lambda}^{m,l}$ and torsion $\cal T$ is introduced as
$${\cal T} = d{\chi} + [{\Gamma}\land \chi \} = {\cal T}^{\underline \alpha}_{\cdot
 \beta \gamma} e_{\underline \alpha} du^{\beta} \land du^{\gamma} ,$$
$ {\cal T}^{\underline \alpha}_{\cdot \beta \gamma} = {A^{\underline \alpha}}_{\alpha}
{T^{\alpha}}_{\cdot \beta \gamma} $ are defined by the components of the torsion
ds--tensor (38).
\par
By using metric G (35) on sv--bundle $\cal E$ we can define the dual (Hodge)
operator ${\ast}_G : {\overline \Lambda}^{q,s} ({\cal E}) \to
{\overline \Lambda}^{n-q, k-s} ({\cal E} ) $ for forms with values in LS--algebras
on ${\cal E}$ (see details, for instance, in [52]), where ${\overline \Lambda}^
{q,s}({\cal E})$ denotes the s--algebra of exterior (q,s)--forms on $\cal E .$
\par
Let operator ${\ast}^{-1}_G$ be the inverse to
operator $\ast$ and ${\hat \delta}_G$ be the adjoint to
the absolute derivation d (associated to the scalar product for s--forms) specified
for (r,s)--forms as $${\delta}_G = {(-1)}^{r+s} {\ast}^{-1}_G \circ d \circ
{\ast}_G .$$
Both introduced operators act in the space of LS--algebra--valued forms as
$$ {\ast}_G ( I_{\hat \alpha} \otimes {\phi}^{\hat \alpha} ) =
I_{\hat \alpha} \otimes ({\ast}_G {\phi}^{\hat \alpha})$$
and
$${\delta}_G ( I_{\hat \alpha} \otimes {\phi}^{\hat \alpha} ) =
I_{\hat \alpha} \otimes {\delta}_G {\phi}^{\hat \alpha} .$$
If the supersymmetric variant of the Killing form for the structural s--group of a
s--bundle into consideration is degenerate as a s--matrix (for instance, this
holds for s--bundle ${\cal A}N({\cal E})$ ) we use an auxiliary nondegenerate
bilinear s--form in order to define formally a metric structure ${G_{\cal A}}$
in the total space of the s--bundle. In this case we can introduce operator
${\delta}_{\cal E}$ acting in the total space and define operator ${\Delta}
\doteq {\hat H} \circ {\delta}_{\cal A} ,$ where ${\hat H}$ is the operator
of horizontal projection. After $\hat H$--projection we shall not have dependence
on components of auxiliary bilinear forms.
\par
Methods of abstract geometric calculus, by using operators $ {{\ast}_G},
{{\ast}_{\cal A}}, {{\delta}_G}, {{\delta}_{\cal A}}$ and ${\Delta} ,$ are
illustrated, for instance, in [54-57] for locally isotropic, and in [32]
for locally anisotropic, spaces. Because on superspaces these operators act in a
similar manner we omit tedious intermediate calculations and present the
final necessary results. For ${\Delta}{\overline B}$ one computers
$${\Delta}{\overline {\cal B}} = ( {\Delta}{\cal B} , {\cal {R\tau}} + {\cal R}i) , $$
where ${\cal {R\tau}} = {\delta}_G {\cal J} + {\ast}^{-1}_G [{\Gamma},{\ast}
{\cal J} \} $ and
\begin{equation}
{\cal R}i  = {{\ast}^{-1}_G} [ {\chi}, {{\ast}_G} {\cal R} \} =
{(-1)}^{n+k+l+m} {{\cal R}_{\alpha \mu}} G^{\alpha {\hat \alpha}}
 {e_{\hat \alpha}}
{\delta u^{\mu}} . \label {f50}
\end{equation}
Form ${\cal R}i$ from (50) is locally constructed by using the components of
the Ricci ds--tensor (40) as it follows from the decomposition with respect to a
locally adapted basis ${\delta u^{\alpha}}$ (7).
\par
Equations
\begin{equation}
{\Delta}{\overline {\cal B}} = 0 \label {f51}
\end{equation}
are equivalent to the geometric form of Yang--Mills equations for the connection
${\overline {\Gamma}}$ (see (49)).
In [55--57] it is proved that such gauge equations coincide with the vacuum
Einstein equations if as components of connection form (47) are used the usual
Christoffel symbols. For spaces with local anisotropy the torsion of
a metric d--connection in general is not vanishing and we have to introduce the source 1--form
in the right part of (51) even gravitational interactions with matter fields
are not considered [32].
\par
Let us consider the locally anisotropic supersymmetric matter source
${\overline {\cal J}}$ constructed by using the same formulas as for
${\Delta}{\overline {\cal B}}$ when instead of ${\cal R}_{\alpha \beta}$ from (50)
is taken  ${{\kappa}_1} ({\cal J}_{\alpha \beta} - {\frac {1}{2}} G_{\alpha \beta} {\cal J}) -
 {\lambda} (G_{\alpha \beta} - {\frac {1}{2}} G_{\alpha \beta}
 G_{\tau}^{\cdot \tau}) .$
By straightforward calculations we can  verify that Yang--Mills equations
\begin{equation}
{\Delta}{\overline {\cal B}} = {\overline {\cal J}}   \label {f52}
\end{equation}
for torsionless connection ${\overline {\Gamma}} = ({\Gamma} , {\chi})$ in
s-bundle ${\cal A}N({\cal E})$ are equivalent to Einstein equations (42)
on sv--bundle $\cal E .$
 But such types of gauge like la-supergravitational
equations, completed with algebraic equations for torsion and s--spin source,
 are not variational in the total space of the s--bundle ${\cal A}L {({\cal E})}.$
This is a consequence of the mentioned degeneration of the Killing form for
the affine structural group [55,56] which also holds for our la-supersymmetric
generalization. We point out that   we have introduced equations (52) in a "pure"
geometric manner by using operators ${\ast},{\quad}{\delta}$ and
horizontal projection ${\hat H} .$
\par
We end this section by emphasizing that
to construct a variational gauge like supersymmetric la--gravitational model
is possible, for instance, by considering a minimal extension of the gauge
s--group $AF^{m,l}_{n,k} ({\Lambda})$ to the de Sitter s--group $S^{m,l}_{n,k}
({\Lambda}) = SO^{m,l}_{n,k} ({\Lambda}) ,$ acting on space ${\Lambda}^{m,l}_{n,k}
\oplus {\cal R} ,$ and formulating a nonlinear version of de Sitter gauge
s--gravity  (see [57] for locally isotropic gauge gravity and [32] for a locally
anisotropic variant). Such s--gravitational models will  be analyzed in details
in [58].

\section{DISCUSSION AND CONCLUSIONS}

In this paper we have formulated the theory of nonlinear and distinguished
connections in sv--bundles which is a framework for developing supersymmetric
models of fundamental physical interactions on la-superspaces. Our approach
has the advantage of making manifest the relevant structures of supersymmetric
theories with local anisotropy and putting great emphasis on the analogy with both
usual locally isotropic supersymmetric gravitational models and locally anisotropic
gravitational theory on vector bundles provided with compatible nonlinear and
distinguished linear connections and metric structures.
\par
The proposed supersymmetric differential geometric techniques allows us a
rigorous mathematical study and analysis of physical consequences of various
variants of supergravitational theories (developed in a manner similar to the
Einstein theory, or in a gauge like form). As two examples we have considered in
details two models of locally anisotropic supergravity which have been chosen
to be equivalent in order to illustrate the efficience and particularities of
applications of our formalism in supersymmetric theories of la--gravity.
\par
We emphasize that there are a number of arguments for taking into account effects of possible local
anisotropy of both the space--time  and fundamental interactions. For example,
it's well known the result that a selfconsistent description of radiational
processes in classical field theories requiers adding of higher derivation terms
(in classical electrodynamics radiation is modelated by introducing a corresponding
term proportional to the third derivation on time of coordinates). A very
important argument for developing quantum field models on the tangent bundle
is the unclosed character of quantum electrodynamics. The renormalized amplitudes
in the framework of this theory tend to $\infty$ with values of momenta $p\to
\infty.$ To avoid this problem one introduces additional suppositions, modifications of
fundamental principles and extensions of the theory, which are less motivated
from physical point of view. Similar constructions, but more sophisticated, are
in order for modelling of radiational dissipation in all variants of classical
and quantum (super)gravity and (supersymmetric) quantum field theories with
higher derivations. It is quite possible that the Early Universe was in a state
with local anisotropy caused by fluctuations of quantum space-time "foam".
\par
The above mentioned points to the necessity to extend the geometric background
of some models of classical and quantum field interactions if a careful
analysis of physical processes with  non--negligible beak reaction, quantum and
statistical fluctuations, turbulence, random dislocations and disclinations in
continuous media and so on.

\acknowledgments

The author would like express his gratitude to officials of the Romanian
Ministry of Research and Technology for their support of investigations on
theoretical physics in the Republic of Moldova and to Academician R.Miron
and Professor M.Anastasiei for useful discussions on generalized Lagrange
geometry and locally anisotropic gravity.

\end{document}